\documentclass[aps,twocolumn,superscriptaddress]{revtex4-2}
\usepackage[T1]{fontenc}
\usepackage{graphicx}
\usepackage{amssymb,amsmath}
\usepackage{braket}
\usepackage{placeins}
\usepackage{tabularx}
\usepackage{siunitx}
\usepackage{physics}
\AtBeginDocument{\RenewCommandCopy\qty\SI}
\usepackage[colorlinks=true, citecolor={blue!80!black}, urlcolor={blue!80!black}, linkcolor = {blue!80!black}]{hyperref}
\usepackage{pifont}
\usepackage{mathtools}
\renewcommand{\vec}[1]{\underline{#1}}
\usepackage{dcolumn}   
\usepackage{array}
\usepackage{placeins}
\makeatletter

\newcommand{\Sb}{^{123}\rm Sb}

\usepackage{xr}

\usepackage{xcolor}
\usepackage{soul}

\newcommand{\beginsupplement}{%
        \setcounter{table}{0}
        \renewcommand{\thetable}{S\arabic{table}}%
        \setcounter{figure}{0}
        \renewcommand{\thefigure}{S\arabic{figure}}%
        \setcounter{section}{0}
        \renewcommand*{\thesection}{S\arabic{section}}%
        \setcounter{equation}{0}
        \renewcommand{\theequation}{S\arabic{equation}}%
     }

\newcommand{\nocontentsline}[3]{}
\newcommand{\tocless}[2]{\bgroup\let\addcontentsline=\nocontentsline#1{#2}\egroup}
\makeatletter
\def\maketitle{
\@author@finish
\title@column\titleblock@produce
\suppressfloats[t]}
\makeatother

\usepackage{xpatch}

\makeatletter
\patchcmd{\@ssect@ltx}
    {\addcontentsline{toc}{#1}{\protect\numberline{}#8}}
    {}
    {}
    {}
\makeatother

\begin{document}

\title{Certifying the quantumness of a nuclear spin qudit through its uniform precession}

\author{Arjen Vaartjes}
\affiliation{School of Electrical Engineering and Telecommunications, UNSW Sydney, Sydney, NSW 2052, Australia}
\affiliation{ARC Centre of Excellence for Quantum Computation and Communication Technology}
\author{Martin Nurizzo}
\affiliation{School of Electrical Engineering and Telecommunications, UNSW Sydney, Sydney, NSW 2052, Australia}
\affiliation{ARC Centre of Excellence for Quantum Computation and Communication Technology}
\author{Lin Htoo Zaw}
\affiliation{Centre for Quantum Technologies, National University of Singapore, 3 Science Drive 2, Singapore 117543}
\author{Benjamin Wilhelm}
\affiliation{School of Electrical Engineering and Telecommunications, UNSW Sydney, Sydney, NSW 2052, Australia}
\affiliation{ARC Centre of Excellence for Quantum Computation and Communication Technology}
\author{Xi Yu}
\affiliation{School of Electrical Engineering and Telecommunications, UNSW Sydney, Sydney, NSW 2052, Australia}
\affiliation{ARC Centre of Excellence for Quantum Computation and Communication Technology}
\author{Danielle Holmes}
\author{Daniel Schwienbacher}
\affiliation{School of Electrical Engineering and Telecommunications, UNSW Sydney, Sydney, NSW 2052, Australia}
\affiliation{ARC Centre of Excellence for Quantum Computation and Communication Technology}
\author{Anders Kringh{\o}j}
\affiliation{School of Electrical Engineering and Telecommunications, UNSW Sydney, Sydney, NSW 2052, Australia}
\affiliation{ARC Centre of Excellence for Quantum Computation and Communication Technology}

\author{Mark R. van Blankenstein}
\affiliation{School of Electrical Engineering and Telecommunications, UNSW Sydney, Sydney, NSW 2052, Australia}
\affiliation{ARC Centre of Excellence for Quantum Computation and Communication Technology}

\author{Alexander M. Jakob}
\affiliation{School of Physics, University of Melbourne, Melbourne, VIC 3010, Australia}
\affiliation{ARC Centre of Excellence for Quantum Computation and Communication Technology}

\author{Fay E. Hudson}
\affiliation{School of
Electrical Engineering and Telecommunications, UNSW Sydney, Sydney, NSW 2052, Australia}
\affiliation{Diraq Pty. Ltd., Sydney, NSW, Australia}

\author{Kohei M. Itoh}

\affiliation{School of Fundamental Science and Technology, Keio University, Kohoku-ku, Yokohama, Japan}

\author{Riley J. Murray}
\affiliation{Quantum Performance Laboratory, Sandia National Laboratories, Livermore, CA 94550, USA}

\author{Robin Blume-Kohout}
\affiliation{Quantum Performance Laboratory, Sandia National Laboratories, Albuquerque, NM 87185, USA}

\author{Namit Anand}

\affiliation{Quantum Artificial Intelligence Laboratory (QuAIL), NASA Ames Research Center, Moffett Field, CA, 94035, USA}
\affiliation{KBR, Inc., 601 Jefferson St., Houston, TX 77002, USA}

\author{Andrew S. Dzurak}
\affiliation{School of
Electrical Engineering and Telecommunications, UNSW Sydney, Sydney, NSW 2052, Australia}
\affiliation{Diraq Pty. Ltd., Sydney, NSW, Australia}

\author{David N. Jamieson}

\affiliation{School of Physics, University of Melbourne, Melbourne, VIC 3010, Australia}
\affiliation{ARC Centre of Excellence for Quantum Computation and Communication Technology}

\author{Valerio Scarani}
\affiliation{Centre for Quantum Technologies, National University of Singapore, 3 Science Drive 2, Singapore 117543}
\affiliation{Department of Physics, National University of Singapore, 2 Science Drive 3, Singapore 117542}

\author{Andrea Morello}
\thanks{Corresponding author; e-mail: \url{a.morello@unsw.edu.au}}
\affiliation{School of Electrical Engineering and Telecommunications, UNSW Sydney, Sydney, NSW 2052, Australia}
\affiliation{ARC Centre of Excellence for Quantum Computation and Communication Technology}


\begin{abstract}
Spin precession is a textbook example of dynamics of a quantum system that exactly mimics its classical counterpart. Here we challenge this view by certifying the quantumness of exotic states of a nuclear spin through its uniform precession. The key to this result is measuring the positivity, instead of the expectation value, of the $x$-projection of the precessing spin, and using a spin $>1/2$ qudit, that is not restricted to semi-classical spin coherent states. The experiment is performed on a single spin-7/2 $\Sb$ nucleus, implanted in a silicon nanoelectronic device, amenable to high-fidelity preparation, control, and projective single-shot readout. Using Schr\"{o}dinger cat states and other bespoke states of the nucleus, we violate the classical bound by 19 standard deviations, proving that no classical probability distribution can explain the statistic of this spin precession, and highlighting our ability to prepare quantum resource states with high fidelity in a single atomic-scale qudit.

\end{abstract}

\maketitle

\tocless\section{Introduction}

Drawing the boundary between classical and quantum behavior has been a long-standing challenge, especially for single-particle systems. 
The spin-1/2 particle, a textbook example of a quantum system, illustrates this difficulty. 
Undergraduate physics students learn that the spin of an electron is quantized: the Stern-Gerlach experiment shows that, upon measuring the orientation of the spin, only one of two discrete possibilities can be observed. However, they soon after learn about the Ehrenfest theorem, which states that the time dependence of the expectation values of dynamical variables are identical for classical and quantum systems with linear Hamiltonians \cite{Ehrenfest1927}. Magnetic resonance textbooks invariably start by showing that the quantum mechanical treatment of a precessing spin yields the same result as the precession of a classical angular momentum \cite{Slichter1990}, leaving many students with the impression that there is nothing quantum about spin precession. 

The key point here -- seldom made explicit by standard textbooks, but crucial to our discussion -- is that the classical-like behaviour of spin-1/2 particles stems from their limited Hilbert space dimension ($d=2$), which can only accommodate semi-classical spin-coherent states \cite{arecchi1972atomic,Kitagawa1993}, and imposes a statistical behavior explainable by noncontextual hidden-variable models \cite{Bell1964_hiddenvariables}. Therefore, higher spins with $d \geq 3$ are required to observe quantum behaviours that cannot be mimicked by classical analogues \cite{Gleason1957, Kochen1967, David1993}.


Even using physical systems with enough complexity and Hilbert space dimension to support nonclassical states, detecting their quantunmess is a delicate task. Conventional approaches have typically relied on measuring correlations between observables using, for instance, Bell nonlocality tests \cite{Bell1964}, Kochen--Specker noncontextuality tests \cite{Kochen1967}, and Leggett--Garg inequality tests \cite{LeggettGarg1985}. However these methods have limitations. Bell nonlocality only probes multipartite notions of quantumness, and thus cannot test the nonclassicality of a single particle. Single-particle tests like Kochen--Specker noncontextuality and Leggett--Garg inequalities require simultaneous or sequential measurements, together with assumptions that are virtually untestable (e.g.~two measurements commute exactly \cite{BK2004}, the system has no classical memory \cite{Kleinmann_2011}, a classical system was measured non-invasively \cite{WildeMizel2012}). Other proposed indicators of nonclassicality like Wigner negativity \cite{KenfackZyczkowski2004} not only require full or partial tomography, but might not have a clear interpretation in finite-dimensional spin systems \cite{Davis2021}. Alternative approaches include evolving the system with higher-order Hamiltonians thought to be needed to overcome the classicality of linear dynamics \cite{Kitagawa1993}.

Here we present the first experimental demonstration of a protocol to detect the quantumness of a spin state via measurements of its precession. Inspired by Tsirelson's work on harmonic oscillators \cite{Tsirelson2006} and later extended to spin systems by \citet{Zaw2022}, this protocol challenges conventional wisdom by revealing clear quantum signatures in a 
system undergoing uniform precession, despite Ehrenfest's theorem suggesting classical-like average behaviour.
The success of the protocol formally disproves the existence of a classical joint probability distribution that could explain the observed statistics, which is a key indicator of nonclassicality \cite{Aravinda2017}. 

The protocol is remarkably simple to implement: it operates under the sole assumption that the measured observable undergoes a uniform precession, i.e. at constant angular velocity, in some suitably defined reference frame. This can readily be implemented in qudits by utilizing recently demonstrated spin-coherent control techniques \cite{Yu2024, Champion2024, Roy2024}. Furthermore, no simultaneous or sequential measurements are required and only $2I$ measurements are required for a spin $I$ system, compared to the $\sim I^2$ required for tomography \cite{Hofmann2004}.

In this work, we implement the precession protocol on a single spin-7/2 $\Sb$ nucleus implanted within a silicon nano-electronic device. We initialize non-classical states and
observe significant violations of the classical bound in the protocol measurements. These violations are possible due to the high fidelity of our prepared states, which we 
verify with quantum state tomography. Furthermore, we also achieve a score close to the general theory-independent maximum possible in a spin-3/2 subspace, which rules out alternatives to quantum theory that cannot reach the same value \cite{Zaw2024}. We shall first present the protocol and its implementation, and leave for the end the discussion on why it works despite widely-held beliefs on the implications of the Ehrenfest theorem, or textbook proofs that spin precession follows the classical dynamics \cite{Slichter1990}. \\\\
\begin{figure*}
    \centering
    \includegraphics{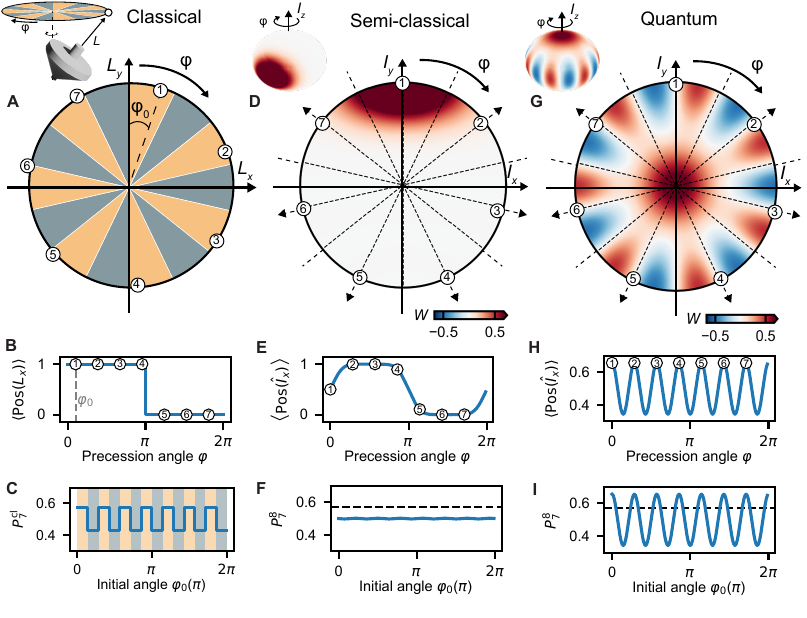}
    \caption{\textbf{Quantumness detection protocol for uniformly precessing 
    spin in classical, semi-classical and quantum states.} 
    (\textbf{A}) Phase space representation of the uniform precession protocol for a classical precessing gyroscope. A precession period is divided into $K$ equal angles ($K=7$ is depicted). Yellow (grey) regions indicate the initial angles $\varphi_0$ that maximize (minimize) the classical score.
    (\textbf{B}) Expectation value of the positivity of $L_x$, as a function of the precession angle $\varphi$. The blue dots indicate a possible set of $K$ equidistantly sampled angles in one precession period. 
    (\textbf{C}) Classical score of the protocol $P_{7}^{cl}$ for $K=7$, as a function of the initial angular shift $\varphi_0$. $P_{K}^{cl}$ is the mean over the $K$ sampled points. The yellow (grey) regions correspond with the yellow (grey) areas in panel A, and indicate the initial angles for which the classical score is maximum (minimum). For $K=7$, the maximum classical score is $P_{K}^{cl}=4/7=0.571$. 
    (\textbf{D}) The Wigner function of a spin-7/2 spin coherent state oriented along the y-axis (inset), with its polar projection for polar angles $\theta \in [0, \pi/2]$. The Wigner function undergoes a rotation around the z-axis. The rotation is again split up in $K$ equal angles.
    (\textbf{E}) Expectation value of the positivity of the $\hat{I}_x$ operator and a possible set of $K=7$ equidistant angles.
    (\textbf{F}) The quantum score $P^8_7$ as a function of the initial angle. Note that the semi-classical spin coherent state does not violate the classical bound (dashed line).
    (\textbf{G}) The Wigner function of a spin-7/2 cat state (inset) and its polar projection for $\theta \in [0, \pi/2]$. (\textbf{H}) The expectation value of ${\rm Pos}(\hat{I}_x$) shows $d-1=7$ oscillation periods, corresponding to the interference pattern along $\varphi$ of the Wigner function. The numbers indicate the set of $K=7$ angles that maximize the quantum score $P^8_7$. 
    (\textbf{I}) $P^8_7$ as a function of the initial shift $\varphi_0$. The maximum quantum score $P^8_7=0.656$ violates the classical bound (dashed line).}
    \label{fig:1}
\end{figure*}

\tocless\section{Results}

\tocless\subsection{Uniform precession protocol}

\tocless\subsubsection{Classical protocol}

To illustrate the classical variant of the protocol, we consider the angular momentum $\vec{L}=\{L_x,L_y,L_z\}$ of a classical gyroscope with a constant precession frequency $\omega$, and a precession angle defined as $\varphi (t) = \omega t + \varphi_0$ (see Fig.~\ref{fig:1}A), where $\varphi_0$ denotes the initial offset angle. The coordinates $L_x$ and $L_y$ of the gyroscope evolve according to the equations of motion:

\begin{equation}
    \begin{gathered} 
    L_x(t) = L_x(0) \cos \varphi (t) + L_y(0) \sin \varphi (t), \\
    L_y(t) = L_y(0) \cos \varphi (t) + L_x(0) \sin \varphi (t).
    \end{gathered}
    \label{eq:motion_classical}
\end{equation}

The protocol involves segmenting one precession period into $K$ equally separated times ($K=7$ in Fig~\ref{fig:1}A). At each of these, we measure the positivity of the $L_x$-coordinate (Fig.~\ref{fig:1}B):

\begin{equation}
    \textrm{Pos}(L_x) = 
    \begin{cases}
        1 & \text{if } L_x\geq 0 \\
        0              & \text{otherwise.}
    \end{cases}
\end{equation}  
The classical score $P_K^{\rm c}=1/K\sum_{k=1}^K {\rm Pos}(L_x)_k$ is defined as the expectation value of $\textrm{Pos}(L_x)$ over one precession period divided into K equal time intervals. In the example, out of 7 intervals, the x-coordinate is positive 4 times, resulting in a classical score of $P^{\rm c}_7=4/7$.

The score $P_K^{\rm c}$ is dependent on the initial state parameterized by $\varphi_0$, i.e., $P_7^{\rm cl}$ is either 3/7 when the state starts in a grey region or 4/7 when it starts in a yellow region, as depicted in Fig.~\ref{fig:1}C. We define the classical bound $\mathbf{P^{\rm c}_K}$ as the maximum classical score, which is obtained by maximizing over all possible initial states. \\\\

\tocless\subsubsection{Quantum protocol} 

For the quantum version of the protocol, we consider the spin vector $\Vec{\hat{I}} = \{\hat{I}_x, \hat{I}_y, \hat{I}_z\}$, as a quantum analogue to classical angular momentum. 
Under a constant $\hat{I}_z$ term in the Hamiltonian, the spin precesses uniformly in the $\hat{I}_x$-$\hat{I}_y$ plane according to the same equations of motion as the classical case (Eq.\ref{eq:motion_classical}), with $L_x \mapsto \hat{I}_x$ and $L_y \mapsto \hat{I}_y$.



Furthermore, the probability of detecting a positive component of $\hat{I}_x$ becomes an operator, i.e. ${\rm Pos}(x) \mapsto {\rm Pos}(\hat{I}_x)$, defined as:

\begin{equation}
    {\rm Pos}(\hat{I}_x) = \frac{1}{2}[\mathbb{I} + {\rm sgn}(\hat{I}_x)],
\end{equation}
where $\mathbb{I}$ is the identity matrix and the sign operator is defined as
\begin{equation}
    {\rm sgn}(\hat{I}_x) = \sum_{m=-I}^I{\rm sgn}(m)\ket{m_x}\bra{m_x}, 
\end{equation}
where $\ket{m_x}$ are the eigenstates of $\hat{I}_x$.
Then for a pure quantum state $\ket{\psi}$ in a $d$-dimensional Hilbert space, the quantum score $P^d_K$ is determined by measuring the expectation value $\big<{\rm Pos}(\hat{I}_x)\big>$ at $K$ separate times during one precession period around the $\hat{I}_z$ axis:
\begin{equation}
    P^d_K = \frac{1}{K}\sum_{k=1}^K\bra{\psi_k} {\rm Pos}(\hat{I}_x) \ket{\psi_k},
\end{equation}
where $\ket{\psi_k} = e^{i (2\pi k / K) \hat{I}_z} \ket{\psi}$ is the rotated initial state. The maximum quantum score is $\mathbf{P_k^d}$ achieved by maximizing over all initial states $\ket{\psi}$.
\\\\

Only a subset of quantum states break the classical bound \cite{Zaw2024}. We illustrate this in Fig.~\ref{fig:1}D-I with two examples on an 8-dimensional spin-7/2 system: the semi-classical spin-coherent state and the quantum spin-cat state.

For the spin-coherent state (Fig.~\ref{fig:1}D), the Wigner quasi-probability distribution is concentrated in one spot of the phase space, qualitatively resembling classical behavior. However, unlike the classical case for which the probability distribution is an infinitesimally small point, its Wigner function spreads out as a Gaussian distribution due to the uncertainty principle ($\delta \hat{I}_x\delta \hat{I}_y \geq |\hat{I}_z|/2$). Despite being a pure quantum state, this state does not break the classical bound. For this state, $\big<\mathrm{Pos}(\hat{I}_x)\big>$ (Fig.\ref{fig:1}E) appears as a smoothed version of the classical case (Fig.\ref{fig:1}B), with the phase space blob essentially acting as a Gaussian filter in the $\varphi$ direction.

In contrast, the quantum spin-cat state, defined as the superposition of two maximally separated spin coherent states \cite{Gupta2024, Yu2024} ($\ket{\rm cat_{7/2}}_z = \left( \ket{7/2} \pm \ket{-7/2}\right) / \sqrt{2}$), exhibits non-classical behavior. Its Wigner function is not concentrated in a single region but is spread out over multiple spots in the phase space, showing a distinct interference pattern along the $\varphi$ direction. The expectation value for the positivity operator $\big<\mathrm{Pos}(\hat{I}_x)\big>$ displays a corresponding oscillation pattern with 7 peaks and 7 valleys. When the $K=7$ equally separated angles are exactly matching the peaks of the oscillation, the quantum score $P_7^8$ is maximized and the classical bound is violated, certifying the quantumness of the spin-cat state. \\\\

\tocless\subsection{Experimental Realization}

We perform the uniform precession protocol on a single ionized $\Sb^{+}$ donor implanted into a silicon chip \cite{Asaad2020}. The nuclear spin $I=7/2$ gives rise to an 8-dimensional Hilbert space, described by the following static spin Hamiltonian in frequency units: 
\begin{equation}
\label{eq:Hamiltonian}
\hat{\mathcal{H}}_{D^{+}} = -\gamma_{\rm n}B_{0}\hat{I}_z+\sum_{\mathclap{\alpha,\beta\in\{x,y,z\}}} Q_\mathrm{\alpha\beta} \hat{I}_\mathrm{\alpha}\hat{I}_\mathrm{\beta},
\end{equation}
where $B_0 =1.384$~T is a static magnetic field, $\gamma_{\rm n}=5.55$~MHz/T is the nuclear gyromagnetic ratio, $\hat{I}_{\{x, y, z\}}$ are the nuclear spin operators,
and $Q_\mathrm{\alpha\beta}$ is the quadrupole strength, which allows for individual addressability \cite{Fernandez2024, Asaad2020}.
The coupling between the 8 different eigenstates is generated on demand by oscillating magnetic fields created via an on-chip antenna. This interaction can induce Nuclear Magnetic Resonance (NMR) transitions and is described by the following interaction Hamiltonian:
\begin{equation}
  \hat{\mathcal{H}}_\mathrm{1}(t) = -\gamma_\mathrm{n} \hat{I}_x \sum_{i=0}^{2I-1}\cos(2\pi f_i t + \phi_i)B_{1,i}(t),
  \label{eq:H_1}
\end{equation}
where $f_k$ are the NMR frequencies ($f_0 = \langle -7/2|\hat{\mathcal{H}}_{D^{+}}| -7/2\rangle - \langle -5/2|\hat{\mathcal{H}}_{D^{+}}| -5/2\rangle$, etc.), $B_{1,i}(t)$ the oscillating magnetic field amplitudes, and $\phi_i$ the phases. The dynamics of the qudit is conveniently described in a Generalized Rotating Frame (GRF) \cite{Leuenberger2003, Yu2024} 
which changes the basis to the rotating eigenstates $\ket{m'_z} = \exp(-i 2 \pi E_{m} t / \hbar)\ket{m_{z}}$, where $E_{m}$ are the eigenenergies of the static Hamiltonian \cite{Neeley2009}. In this basis, the Hamiltonian is transformed into:
\begin{equation}
\hat{\mathcal{H}}_{\mathrm{G}} = 
  \left(
  \begin{array}{ccccccc}
  0 & g_0 e^{i\phi_0} & 0 & \cdots & 0 & 0 \\
  g_0^* e^{-i\phi_0} & 0 & g_1 e^{i \phi_1} & \cdots & 0 & 0 \\
  0 & g_1^* e^{-i \phi_1} & 0 & \cdots & 0 & 0 \\
  \vdots & \vdots & \vdots & \ddots & \vdots & \vdots \\
  0 & 0 & 0 & \cdots & 0 & g_6 e^{i\phi_6} \\
  0 & 0 & 0 & \cdots & g_6^* e^{-i\phi_6} & 0 \\
  \end{array}
  \right),
\end{equation}
where $g_i=-\frac{\gamma_n}{4} B_{1,i}(t)\bra{-7/2+i}\hat{I}_x\ket{-7/2+(i+1)}$.

\begin{figure}
    \centering
    \includegraphics{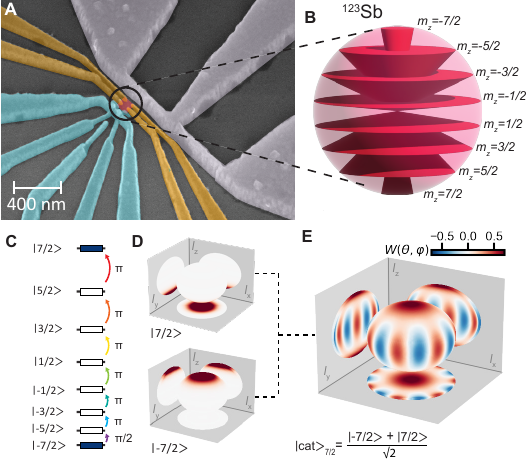}
    \caption{\textbf{Preparation of non-classical states of a $\Sb$ nucleus in a silicon device}.
    (\textbf{A}) Scanning electron micrograph of a device identical to that used in this experiment. The ion-implanted $^{123}$Sb nucleus is located in the donor implant window (red), see section IV.A. The electrostatic donor gates (yellow), single electron transistor (SET) (teal) and microwave antenna (light grey) are shown.
    (\textbf{B}) Spin angular momentum states of the $^{123}$Sb nucleus under an external magnetic field, showing 8 different quantum levels $m_z$ of the nucleus. 
    (\textbf{C}) State initialization process, using ladder operations (Givens rotations) to create a spin cat state: a sequence of $\pi/2$ and $\pi$-pulses (ordered bottom to top) drive transitions between different $\ket{m_I}$ eigenstates of the $\hat{I}_z$-operator. 
    (\textbf{D}) Three-dimensional visualization of the spin-Wigner function $W(\theta, \phi)$ for the spin-coherent eigenstates $\ket{7/2}$ and $\ket{-7/2}$. 
    (\textbf{E}) Spin-Wigner function $W(\theta, \varphi)$ (section SI.E) of the cat state $\ket{-7/2} - \ket{-7/2}/\sqrt{2}$, illustrating the interference fringes characteristic of a non-classical spin-cat state.
    }
    \label{fig:2}
\end{figure}

This Hamiltonian outlines the experimental tools for the precession protocol.
State initialization is performed (see supplementary section SII.B) with two-level ladder operations or Givens rotations \cite{Cybenko2001} with sequentially applied single-tone drives, activating only one $B_{1,i}$ at a time. Figure~\ref{fig:2}C shows the example of preparing the spin-cat state $\ket{\rm cat_{7/2}}_z = \left( \ket{7/2}-\ket{-7/2}\right) / \sqrt{2}$. 
Furthermore, any state can be rigidly rotated by SU(2)-covariant rotations, produced by driving all NMR transitions with tones of equal amplitude $B_{1,i} = B_1$ \cite{Yu2024}. 
Finally, the uniform precession required for the protocol is implemented through virtual selective number arbitrary phase (virtual-SNAP) gates. When applied with a specific set of phases, these gates generate virtual $\hat{I}_z$ rotations. Here, these rotations are achieved by updating the GRF, through equal phase shifts of all 7 reference clocks by $\phi$ (see supplementary section SII.C for a detailed description). \\\\

\tocless\subsection{Breaking the classical bound with spin cat states}

\begin{figure*}
    \centering
    \includegraphics{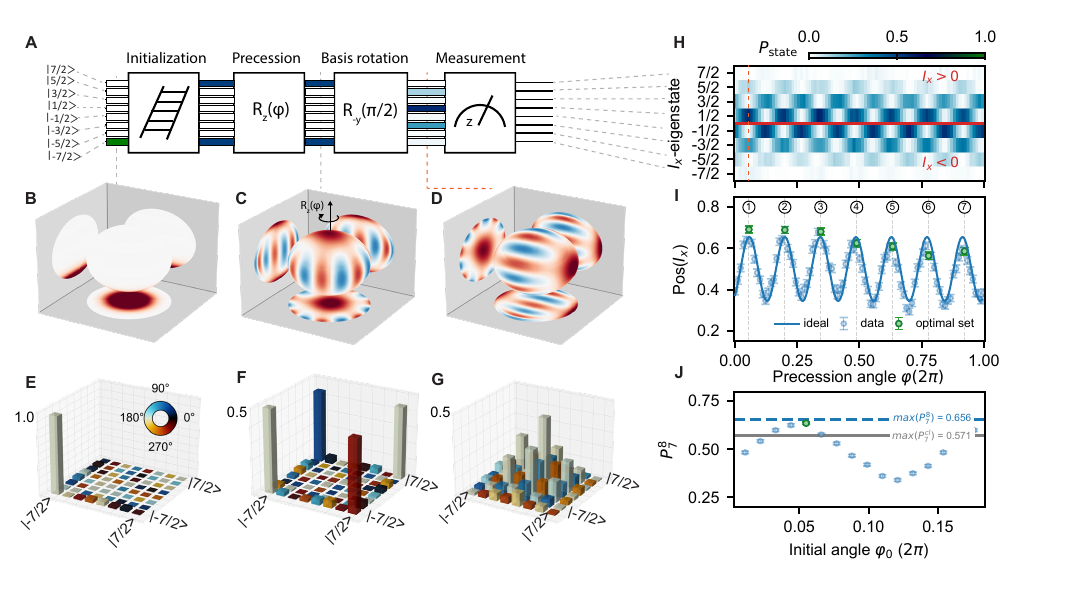}
    \caption{\textbf{Experimental realization of the protocol for detecting quantumness in uniform precession.} (\textbf{A}) The sequence of operations in the protocol: initialization, precession, basis rotation, and measurement. Initialization is achieved through ladder operations to prepare the nucleus in a specific state. During precession, the state rotates around the $ z $-axis by an angle $\varphi$. Basis rotation is performed using an SU(2) $ \pi/2 $ rotation around the $-y$-axis, followed by measurement in the $ z $-basis. The bar colors in the pulse sequence represent state probabilities at each protocol step, with the same color scale as panel \textbf{H}, and the colors before the measurement correspond to the line cut (orange dashed line in panel \textbf{H}). (\textbf{B}-\textbf{D}) Experimentally reconstructed Wigner functions show the state evolution throughout the protocol, obtained via quantum state tomography (see Supplementary Information III.A). (\textbf{E}-\textbf{G}) Density matrices corresponding to the Wigner plots. (\textbf{H}) Populations of the $ \hat{I}_x $-eigenstates $ |m_x \rangle $ as a function of the precession angle $ \varphi $. The populations are inferred from the probabilities of measuring the $\ket{m_z}$ states after the basis rotation. The red horizontal line separates positive from negative eigenstates. (\textbf{I}) The expectation value of the positivity operator $ \text{Pos}(\hat{I}_x) $ (blue data points) shows oscillations in $ \varphi $, closely matching the simulated ideal behavior (blue solid line) of a spin-cat state. The numbered green data points indicate the optimal set of $K=7$ equally separated points that maximize the measured quantum score in the protocol. (\textbf{J}) The highest measured quantum score for $ d = 8 $ and $ K = 7 $, $ P_7^8 = 0.636(7)$, approaches the theoretical maximum quantum score $ \mathbf{P_7^8} = 0.656 $ and exceeds the classical bound $\mathbf{P_7^{\text{c}}}=0.571$ by 19 standard deviations. The error bars in (\textbf{I}-\textbf{J}) represent $2\sigma$ confidence intervals.}
    \label{fig:3}
\end{figure*}

\tocless\subsubsection{Pulse sequence}

The protocol is summarized by the following four-step pulse sequence on the ground state $\ket{-7/2}$ (Fig.~\ref{fig:3}A):
\begin{enumerate}
    \item Initialize the desired state using two-level ladder operations. 
    \item Emulate uniform precession around the $\hat{I}_z$-axis by performing an $R_z$ rotation with angle $\varphi$.
    \item Perform a $\pi/2$ SU(2) rotation around the $-y$ axis to change the measurement basis from $\hat{I}_z$ to $\hat{I}_x$.
    \item Measure the state populations in the $\hat{I}_z$-basis.
\end{enumerate}



This sequence allows us to measure $\rm{Pos}(\hat{I}_x)$ at $K$ angles $\varphi_k$ and determine a quantum score $P^d_K$. 

The Schr\"{o}dinger cat state -- a paradigmatic example of a nonclassical state, and a useful resource in quantum error correction \cite{Puri2020, Gross2021, Gross2024} and quantum sensing \cite{Facon2016, Chalopin2018} -- is a particularly suitable state on which to test this protocol.
Figure~\ref{fig:3} demonstrates the breaking of the classical bound using the largest cat state, $\ket{\rm cat_{7/2}}_z$ \cite{Zaw2022}. By selecting $K=7$ equidistant points along the $\varphi$ axis and averaging the values of $\textrm{Pos}(\hat{I}_x)$ at these points, we obtain a quantum score $P^8_7 = 0.636(7)$ (throughout this work, the error represents a $2\sigma$ confidence interval). This is an observed violation of the classical bound by 19 standard deviations, which certifies the quantumness of state we prepared on the $^{123}$Sb nuclear spin.

Such a large violation must mean that we were able to prepare $\ket{\rm cat_{7/2}}_z$ with a high fidelity. Indeed, we can quantitatively lower bound the fidelity with the observed score as $\mathcal{F} \geq 16(2P^8_7-1)/5 = 87\%$ (see Supplementary SI.C), which is a conservative estimate of the fidelity of $\mathcal{F} = 95.3(1.2)\%$ obtained via quantum state tomography (detailed in Supplementary SIII.A).

Tomography also allows us to visualize the Wigner sphere at each step of the process (Fig.~\ref{fig:3}B-D). The plot of the violating state $\ket{\rm cat_{7/2}}_z$ (Fig.~\ref{fig:3}C) shows that it exhibits an interference pattern, which is read out by rotating the axis initially along $\hat{I}_x$ to the $\hat{I}_z$ direction (Fig.~\ref{fig:3}D), thereby equivalent to a measurement in the $x-$basis. 

This interference pattern is not only a typical signature of quantum effects, but also provides a qualitative explanation of the quantum score. Fig.~\ref{fig:3}H is key to understanding the physics behind the protocol. As the cat state rotates by an angle $\varphi$ through the uniform precession, the populations of the $\ket{m_x}$ eigenstates oscillate with $d-1=7$ periods over a $2\pi$ rotation. However, the $I_x$ expectation value remains identically zero at all times: this is what would be measured in a standard precession experiment with inductive detection. Conversely, the positivity $\textrm{Pos}(\hat{I}_x)$, obtained by summing the probabilities of the states with $\hat{I}_x > 0$, is not constant: taking the cut along the orange dashed line in Fig.~\ref{fig:3}H shows that the spin is more likely to be in the $I_x>0$ sector, whereas the pattern reverses shortly thereafter. 

Figure~\ref{fig:3}I shows how to break the classical bound using the $\ket{\rm cat_{7/2}}_z$ state. We sample the precession at the 7 values of $\varphi$ where the oscillation pattern in $\textrm{Pos}(\hat{I}_x)$ is at its maximum. The oscillating pattern itself arises from the quantum interference of the $\ket{\pm 7/2}$ states that are superimposed to form the Schr\"{o}dinger cat states -- a pattern that cannot be present for a classical gyroscope. In fact, obtaining $\mathbf{P}_K^d > \mathbf{P}_K^{\rm c}$ with $K=7$ also certifies our system to be a qudit with dimension $d \geq 8$, as it is known that no violation of the classical bound is possible when $d \leq K$ \cite{Zaw2022}.

\tocless\subsubsection{Breaking the bound with smaller spins}

\begin{figure}
    \centering
    \includegraphics{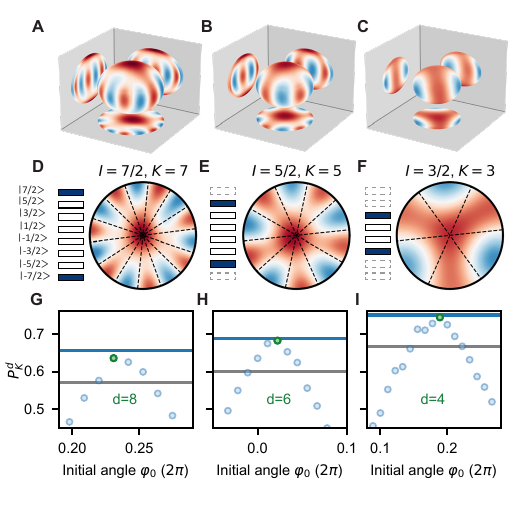}
    \caption{\textbf{Breaking the classical bound in different size subspaces.}
    (\textbf{A}-\textbf{C}) Reconstructed Wigner functions of the spin-cat states in (\textbf{A}) $d=8$, (\textbf{B}) $d=6$, and (\textbf{C}) $d=4$. The size of the sphere reflects the size of the subspace. The Wigner spheres are reconstructed from full $8-$dimensional quantum state tomography (see Supplementary SIII.A). 
    (\textbf{D}-\textbf{F}) Population distribution (bars left) and polar projection (right) of the spin-cat states. (\textbf{D}) The $d=8$ cat state occupies the whole Hilbert space of the $\Sb$ nucleus. (\textbf{E}) The $d=6$ cat state is spanned by the $\ket{\pm 5/2}$ states. (\textbf{F}) The $d=4$ cat state is the superposition of the $\ket{\pm 3/2}$ states.
    (\textbf{G}-\textbf{I}) The measured quantum scores $P^d_k$ (blue datapoints) exceed the classical bound for a small range of initial angles $\varphi_0$. The optimal measured quantum score (green dot) approaches the theoretical maximum (blue horizontal line), and violates the classical bound (grey horizontal line) for all even $d>3$. }
    \label{fig:4}
\end{figure}

To demonstrate the versatility of the protocol, we applied it to smaller dimensional subspaces of the $\Sb$ nuclear spin, showing that we can break the corresponding classical bound for each even dimension as long as $d > 3$. By focusing on even-dimensional subspaces, we effectively emulate smaller fermionic spins, such as the $I=3/2$ spin represented by the subspace spanned by \{${\ket{-3/2}, \ket{-1/2}, \ket{1/2}, \ket{3/2}}$  \}\cite{Neeley2009}.

Adapting the precession protocol to these subspaces requires slight modifications to the initialization, precession, and basis rotation steps in the pulse sequence. In the initialization step, we generate subspace cats for $d=8$, $6$, and $4$ using the ladder initialization method, as described in Fig.\ref{fig:2}C, and restrict the process to the respective subspaces (Fig.\ref{fig:4}D-F). During the uniform precession step, we shift only the relevant $d-1$ clock phases in the subspace by $\varphi$ to achieve a virtual $R^d_{z}(\varphi)$ rotation (detailed in Supplementary II.D). Finally, the basis rotation $R^d_{-y}(\pi/2)$ in the subspace is performed by rescaling the drive amplitudes, ensuring the driving strength, determined by $B_{1,i}$ and the spin-7/2 operator element $\hat{I}^8_x$, matches the spin operator in the subspace $\hat{I}^d_x$ (see Supplementary section SII.D for more details).

Figure~\ref{fig:4} shows the results of the protocol applied to these subspaces. The data illustrates how the protocol successfully breaks the classical bound for each tested even dimension.\\\\

\tocless\subsection{Maximizing the quantum score with unequal intervals}

\begin{figure}[b]
    \centering
    \includegraphics{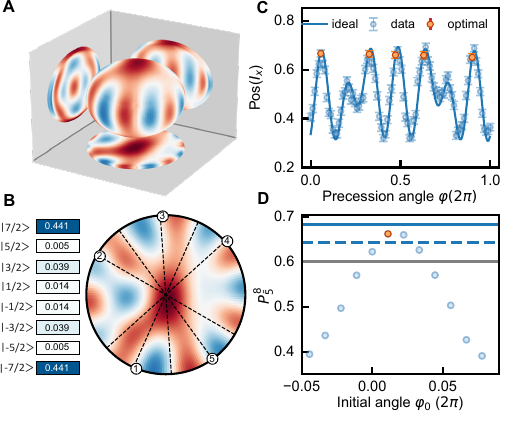}
    \caption{{Breaking the classical bound in the modified protocol, with $\Tilde{K}$ unevenly spaced angles.}
    (\textbf{A}) Spin-Wigner function of the $\ket{P^8_{\Tilde{5}}}$ state.
    (\textbf{B}) Population distribution (left bars) and polar projection of the Wigner function (right), with the numbered dots indicating the unequally spaced angles $\varphi_{\Tilde{k}}$. (\textbf{C}) The expectation value of Pos($\hat{I}_x$) displays five distinct peaks in the data. The orange datapoints shows the optimal set of $\Tilde{K}=5$ unequally separated angles. 
    (\textbf{D}) The quantum score of the modified protocol $P^8_{\Tilde{5}}$ as a function of the initial angle $\varphi_0$, whilst keeping the relative separation between the sampled angles fixed. We measure an increase in the quantum score beyond the maximum score in the original protocol (blue dashed line) $P^8_{\Tilde{5}} = 0.662$, maximizing the gap to the classical score (grey line). The blue solid line indicates the theoretical maximum score. }
    \label{fig:5}
\end{figure}

Finally we perform a slightly modified version of the protocol to maximize the gap between the classical score and the quantum score. In the modified protocol we loosen the restriction that  the $K$ measurement angles have to be equally separated. Instead, the measurement angles $\varphi_{\Tilde{k}}$ follow the rule: $\forall \Tilde{k} : |\varphi_{\Tilde{k}} - 2\pi \Tilde{k}/\Tilde{K}| \leq \pi/(\Tilde{2K})$, where we note the variables in the modified protocol ($\Tilde{k}, \Tilde{K}$) with a tilde. Importantly, the loosened restriction leaves the classical bound unchanged but it allows for larger quantum scores for more complicated states.

For example, the state that maximizes the violation for $d=8$ and $\Tilde{K}=5$ is given by: 

\begin{equation}
    \begin{aligned}
    \ket{P^8_{\Tilde{5}}} &= 0.665 \ket{-7/2} + 0.072 \ket{-5/2} - 0.199 \ket{-3/2} \\ 
    &+ 0.117 \ket{-1/2} - 0.117 \ket{1/2} + 0.199 \ket{3/2} \\
    &+ 0.072 \ket{5/2} - 0.665 \ket{7/2},
    \end{aligned}
\end{equation}
using the set of angles $\{\varphi_k\} = \{-\varphi_1, -\varphi_0, 0, \varphi_0,\varphi_1\}$, where $\varphi_0 = 0.305\pi$ and $\varphi_1 = 0.850\pi$.

The maximally violating states were found by performing gradient descent on $\max P^d_{\tilde{K}}$ over the probing angles $\{\varphi_k\}_k$ for a selection of initial angles that cover the parameter space. Both the derived analytical expression for the gradient and details about the optimisation can be found in Supplementary SI.D. \\\\



\renewcommand{\arraystretch}{1.7}

\begin{table*}[]
\begin{tabularx}{\textwidth}{|>{\hsize=9.4cm}X|X|X|X|X|}
\hline
\multicolumn{5}{|c|}{Spin 7/2 (d=8)} \\ 
\hline
\hline
State & \textbf{$K$ $(\Tilde{K})$} & (max.) $\mathbf{P^{\rm c}_K}$ & (meas.) $P^d_K$ & (max.) $\mathbf{P^{d}_K}$ \\ 
\hline
$\ket{P^8_7}=\frac{1}{\sqrt{2}}\ket{-7/2}-\frac{1}{\sqrt{2}}\ket{7/2}$ & $K=7$ & 0.571 & 0.636(7) & 0.656 \\ 
$\ket{P^8_5}=\frac{1}{\sqrt{2}}\ket{-7/2}-\frac{1}{\sqrt{2}}\ket{3/2}$ & $K=5$ & 0.6 & 0.627(8) & 0.643  \\ 
$\ket{P^8_3}=\frac{\sqrt{7}}{4}\ket{-7/2} -\frac{1}{\sqrt{2}}\ket{-1/2} + \frac{1}{4}\ket{5/2}$ & $K=3$ & 0.667 & 0.695(9) & 0.698  \\ 
$\begin{aligned}
\ket{P^8_{\Tilde{5}}}& =  0.665  \ket{-7/2}  +  0.072 \ket{-5/2}  - 0.199 \ket{-3/2} + 0.117 \ket{-1/2} \\ 
& - 0.117 \ket{+1/2} + 0.199 \ket{+3/2} + 0.072 \ket{+5/2} - 0.665 \ket{+7/2}
\end{aligned}$ & 
$\begin{aligned}
   & \Tilde{K}=5 \\ &\mathrm{(uneven)}  
\end{aligned}$
& 0.6 & 0.662(8) & 0.683  \\ 
$\begin{aligned}
\ket{P^8_{\Tilde{3}}} & = 0.600 \ket{-7/2} - 0.145 \ket{-5/2} - 0.336 \ket{-3/2}- 0.078 \ket{-1/2} \\ 
& + 0.078 \ket{+1/2} + 0.336 \ket{+3/2} + 0.145 \ket{+5/2} - 0.600 \ket{+7/2}
\end{aligned}$ & $\begin{aligned}
   & \Tilde{K}=3 \\ &\mathrm{(uneven)}  
\end{aligned}$
& 0.667 & 0.720(9) & 0.745  \\ 
\hline
\hline
\multicolumn{5}{|c|}{Spin 5/2 (d=6)} \\ 
\hline
\hline
$\ket{P^6_5}=\frac{1}{\sqrt{2}}\ket{-5/2}-\frac{1}{\sqrt{2}}\ket{+5/2}$ & $K=5$ & 0.6 & 0.682(9) & 0.688 \\
$\ket{P^6_3}=\frac{1}{\sqrt{2}}\ket{-5/2}-\frac{1}{\sqrt{2}}\ket{+1/2}$ & $K=3$ & 0.667 & 0.684(10) & 0.698 \\
$\begin{aligned}
\ket{P^6_{\Tilde{3}}} & = 0.645 \ket{-5/2} - 0.119 \ket{-3/2} - 0.264 \ket{-1/2} \\
& - 0.264 \ket{+1/2} - 0.119 \ket{+3/2} + 0.645 \ket{+5/2}
\end{aligned}$ & 
$\begin{aligned}
   & \Tilde{K}=3 \\ &\mathrm{(uneven)}  
\end{aligned}$
& 0.667 & 0.715(10) & 0.746  \\ 
\hline
\hline
\multicolumn{5}{|c|}{Spin 3/2 (d=4)} \\ 
\hline
\hline
$\ket{P^4_3}=\frac{1}{\sqrt{2}}\ket{-3/2}-\frac{1}{\sqrt{2}}\ket{3/2}$ & $K=3$ & 0.667 & 0.744(9) & 0.750 \\
\hline
\end{tabularx}
\label{tab:1}
\caption{Summary of experimental results for different size subspaces $d=$8, 6, and 4. For each $K$ (evenly spaced protocol) or $\Tilde{K}$ (unevenly spaced protocol), we display the optimal state and compare the measured $P^d_k$ with the classical bound $\mathbf{P_K^c}$ and the maximum quantum score $\mathbf{P_K^d}$. For all combinations of $d$ and $K$ ($\Tilde{K}$), we find a violation of the classical bound. The uncertainties in the measured $P_K^d$ represent a 95\% confidence interval.}
\end{table*}

\tocless\section{Discussion}

In this work, we successfully certified quantumness in the time evolution of an 8-dimensional nuclear qudit through its spin precession. 
Our results show a violation of the classical bound for various sized subspaces and measurement intervals, summarized in Tab.~\ref{tab:1} (detailed protocol data and quantum state tomography data available in Supplementary section SIII.C). 


Our observation of quantumness in a spin precession may seem at odds with Ehrenfest's theorem. However, in reality, two conditions must be met for Ehrenfest's theorem to yield classical-like dynamics: the Hamiltonian must be linear in the observables of interest, and the measured observable must have canonical commutation relations with the Hamiltonian \cite{Wheeler1998, hall2013quantum}. In our case, while we indeed generated a linear $\hat{I}_z$ Hamiltonian, we measured the positivity operator ${\rm Pos}(\hat{I}_x)$, which does not obey the commutation relationships with $\hat{I}_z$ that would make it behave classically. This choice of observable allows for a deviation from classical behaviour if the wavefunction of the precessing system is not concentrated in a single region in phase space \cite{hall2013quantum}. Conversely, if one were to measure $\hat{I}_x$ or $\hat{I}_y$ directly -- as is done e.g. in ensemble spin resonance experiments with inductive detection -- no deviation from classical behaviour would be found. Furthermore, in the case of Schr\"{o}dinger cat states as adopted here, all spin expectation values vanish, yielding zero induction signals; the Ehrenfest theorem would provide the uninsightful result $\partial 0/\partial t = 0$. 


From a foundational perspective, the precession protocol complements other tests of quantumness, such as Bell tests \cite{Bell1964}, contextuality experiments \cite{LeggettGarg1985} and tests on the reality of the wavefunction \cite{Ringbauer2015}. Bell tests primarily probe nonlocality, and contextuality tests examine the incompatibility of classical variables with quantum measurements. The precession protocol is more alike the latter in that they both falsify the existence of an underlying ontological model with a classical probability distribution that simply ``\emph{reveals}'' the measured observables. However, they differ in assumptions: the precession protocol assumes that the measured observables follow a certain dynamics, while contextuality experiments assume that the measured observables are compatible and thus commuting.

The measured score of $P_3^4 = 0.744(9)$ obtained in our experiment with the spin-$3/2$ is very close to the general theory-independent bound of $\mathbf{P}_3^G = 0.75$ 
\cite{Zaw2024}. As this bound is saturated by quantum theory, we have effectively ruled out any alternatives to quantum theory that cannot achieve the same score.

The nonclassical states we have efficiently certified through this protocol are of broad interest for quantum information processing \cite{Gross2024,Yu2024,Champion2024,Roy2024}, quantum sensing \cite{Chalopin2018} and quantum foundations \cite{frowis2018macroscopic}. The precession protocol adopted here may become a powerful tool for certification of quantunmess of resource states, because it is fairly economical with respect to the type and number of measurements needed. First, in principle, it only requires measuring the sign of the component of the angular momentum along a certain direction. As such, we would not need to resolve the energy level splitting precisely, as long as the splitting between the positive and negative values are large enough to be distinguished.
Second, the precession protocol in principle only requires $K=7$ measurement directions for a spin-$7/2$, compared to the 45 directions used for full state tomography \cite{Yu2024}. In general, for a spin-$I$ system, only $2I$ measurements of dichotomic observables are required to detect nonclassicality with the precession protocol, compared to $\sim I^2$ measurements of $(2I+1)$--outcome observables to tomograph the state. \\\\

\tocless\section{Methods}

\tocless\subsection{Fabrication}

The qudit system used in this experiment is the nucleus of a $^{123}$Sb donor, introduced into isotopically enriched $^{28}$Si by ion implantation. First, a 900~nm thick epilayer of $^{28}$Si (730~ppm residual $^{29}$Si) was deposited on a natural Si handle wafer. N-type ohmic leads and p-type channel stoppers to prevent leakage currents were formed by thermal diffusion of phosphorus and boron, respectively. A 200~nm thick SiO$_2$ field oxide and an 8~nm thick high quality SiO$_2$ gate oxide were grown in oxidation furnaces. $^{123}$Sb$^+$ ions (18~keV, $5\times10^{11}$~cm$^{-2}$) were implanted at normal incidence through a 90~nm $\times$ 100~nm implant window in a PMMA mask. A rapid thermal anneal at 1000~$^{\circ}$C for 10~s in nitrogen atmosphere was performed to repair the implantation damage and activate the donors. Surface nanoelectronics were fabricated as standard for our qubit devices using three layers of electron beam lithography and aluminium deposition. Each layer is electrically insulated from the others with native Al$_2$O$_3$. Finally, the sample was annealed in forming gas (400~$^{\circ}$C, 15~min, 95$\%$ N$_2$ : 5$\%$ H$_2$) to passivate interface traps. \\\\

\tocless\subsection{Experimental setup}

The device was wire-bonded to a gold-plated PCB and placed in a copper enclosure within a superconducting solenoid generating a \qty{1.384}{\tesla} magnetic field. The setup was cooled to \qty{18}{\milli\kelvin} in a Bluefors BF-LD400 dilution refrigerator. DC voltages were applied on the electrostatic gates with SRS SIM928 sources. The donor gates used for fast pulsing used a room-temperature resistive power combiner. The AC component on those gates were generated using a Quantum Machines OPX+ and were filtered with an \qty{80}{\mega\hertz} low-pass filter. Signals on the DC gates were filtered with a \qty{20}{\hertz} low-pass filter. All filtering occurred at the mixing chamber plate, with graphite-coated coaxial cables. ESR transitions were induced using a Keysight E8267D PSG microwave source (\qty{38.94269}{\giga\hertz} carrier frequency) with single-sideband modulation. The lower-frequency (around \qty{7}{\mega\hertz}) NMR pulses were generated by the OPX+. The ESR and NMR signals were combined before going into the fridge by a Marki Microwave DPX-1721 diplexer. The SET current was amplified by a Femto DLPCA-200, SRS SIM910 before being filtered by an SRS SIM965, and digitized via the OPX+. \\\\

\tocless\subsection{Data and code availability}

All data and code to support the claims in the text can be found in the online repository at \url{https://doi.org/10.5061/dryad.547d7wmj0}. 
Requests for further information and resources should be directed to and will be fulfilled by the lead contact, Andrea Morello (\url{a.morello@unsw.edu.au}) \\\\




\tocless\section{Acknowledgments}
The research was funded by an Australian Research Council Discovery Project (grant no. DP210103769) and the US Army Research Office (contract no. W911NF-23-1-0113).
We acknowledge the facilities, and the scientific and technical assistance provided by the UNSW node of the Australian National Fabrication Facility (ANFF), and the Heavy Ion Accelerators (HIA) nodes at the University of Melbourne and the Australian National University. ANFF and HIA are supported by the Australian Government through the National Collaborative Research Infrastructure Strategy (NCRIS) program. Ion beam facilities employed by D.N.J. and A.M.J. were co-funded by the Australian Research Council Centre of Excellence for Quantum Computation and Communication Technology (Grant No. CE170100012).
X.Y., B.W., M.R.v.B., A.V. acknowledge support from the Sydney Quantum Academy. D.N.J. acknowledges the support of a Royal Society (UK) Wolfson Visiting Fellowship RSWVF/211016.
L.H.Z. and V.S. are supported by the National Research Foundation, Singapore, and A*STAR under its CQT Bridging Grant.
N.A. is a KBR employee working under the Prime Contract No. 80ARC020D0010 with the NASA Ames Research Center and is grateful for the collaborative agreement between NASA and CQC2T. The United States Government retains, and by accepting the article for publication, the publisher acknowledges that the United States Government retains, a nonexclusive, paid-up, irrevocable, worldwide license to publish or reproduce the published form of this work, or allow others to do so, for United States Government purposes.
Sandia National Laboratories is a multi-program laboratory managed and operated by National Technology and Engineering Solutions of Sandia, LLC., a wholly owned subsidiary of Honeywell International, Inc., for the U.S. Department of Energy's National Nuclear Security Administration under contract DE-NA-0003525. All statements of fact, opinion or conclusions contained herein are those of the authors and should not be construed as representing the official views or policies of the U.S. Department of Energy or the U.S. Government. \\\\

\tocless\section{Author contributions}   

Conceptualization: 
A.V., M.N., L.H.Z., B.W, X.Y., D.H., D.S., A.K., M.R.v.B., N.A., V.S., A.M.; 
Investigation:
A.V., M.N.;
Methodology: 
A.V., M.N., L.H.Z., B.W., X.Y., D.H., D.S., A.K., M.R.v.B., R.J.M., R.B-K., N.A.; 
Resources:
D.H., A.M.J., F.E.H., K.M.I., D.N.J., A.M.; 
Software: 
A.V., M.N., B.W., X.Y., M.R.v.B.; 
Writing – original draft:
A.V., M.N., L.H.Z., V.S., A.M.; 
Writing – reviewing and editing:
B.W., X.Y., D.H., D.S., A.K., M.R.v.B., R.B-K., N.A.; 
Supervision:
A.S.D., D.N.J., V.S., A.M.; 
Funding acquisition: 
D.N.J., V.S., A.M.



\begin{thebibliography}{38}
\providecommand{\natexlab}[1]{#1}
\providecommand{\url}[1]{\texttt{#1}}
\expandafter\ifx\csname urlstyle\endcsname\relax
  \providecommand{\doi}[1]{doi: #1}\else
  \providecommand{\doi}{doi: \begingroup \urlstyle{rm}\Url}\fi

\bibitem[Ehrenfest(1927)]{Ehrenfest1927}
P.~Ehrenfest.
\newblock {Bemerkung {\"u}ber die angen{\"a}herte G{\"u}ltigkeit der klassischen Mechanik innerhalb der Quantenmechanik}.
\newblock \emph{Zeitschrift f{\"u}r Physik}, 45\penalty0 (7):\penalty0 455--457, July 1927.
\newblock ISSN 0044-3328.
\newblock \doi{10.1007/BF01329203}.

\bibitem[Slichter(1990)]{Slichter1990}
Charles~P. Slichter.
\newblock \emph{Principles of {{Magnetic Resonance}}}, volume~1 of \emph{Springer {{Series}} in {{Solid-State Sciences}}}.
\newblock Springer, Berlin, Heidelberg, 1990.
\newblock ISBN 978-3-642-08069-2 978-3-662-09441-9.
\newblock \doi{10.1007/978-3-662-09441-9}.

\bibitem[Arecchi et~al.(1972)Arecchi, Courtens, Gilmore, and Thomas]{arecchi1972atomic}
F.~T. Arecchi, Eric Courtens, Robert Gilmore, and Harry Thomas.
\newblock Atomic {{Coherent States}} in {{Quantum Optics}}.
\newblock \emph{Physical Review A}, 6\penalty0 (6):\penalty0 2211--2237, December 1972.
\newblock \doi{10.1103/PhysRevA.6.2211}.

\bibitem[Kitagawa and Ueda(1993)]{Kitagawa1993}
Masahiro Kitagawa and Masahito Ueda.
\newblock Squeezed spin states.
\newblock \emph{Physical Review A}, 47\penalty0 (6):\penalty0 5138--5143, June 1993.
\newblock \doi{10.1103/PhysRevA.47.5138}.

\bibitem[Bell(1966)]{Bell1964_hiddenvariables}
J.~S. Bell.
\newblock On the {{Problem}} of {{Hidden Variables}} in {{Quantum Mechanics}}.
\newblock \emph{Reviews of Modern Physics}, 38\penalty0 (3):\penalty0 447--452, July 1966.
\newblock \doi{10.1103/RevModPhys.38.447}.

\bibitem[Gleason(1957)]{Gleason1957}
Andrew~M. Gleason.
\newblock Measures on the {{Closed Subspaces}} of a {{Hilbert Space}}.
\newblock \emph{Journal of Mathematics and Mechanics}, 6\penalty0 (6):\penalty0 885--893, 1957.
\newblock ISSN 0095-9057.

\bibitem[Kochen and Specker(1967)]{Kochen1967}
Simon Kochen and E.~P. Specker.
\newblock The {{Problem}} of {{Hidden Variables}} in {{Quantum Mechanics}}.
\newblock \emph{Journal of Mathematics and Mechanics}, 17\penalty0 (1):\penalty0 59--87, 1967.
\newblock ISSN 0095-9057.

\bibitem[Mermin(1993)]{David1993}
N.~David Mermin.
\newblock Hidden variables and the two theorems of {{John Bell}}.
\newblock \emph{Reviews of Modern Physics}, 65\penalty0 (3):\penalty0 803--815, July 1993.
\newblock \doi{10.1103/RevModPhys.65.803}.

\bibitem[Bell(1964)]{Bell1964}
J.~S. Bell.
\newblock On the {{Einstein Podolsky Rosen}} paradox.
\newblock \emph{Physics Physique Fizika}, 1\penalty0 (3):\penalty0 195--200, November 1964.
\newblock \doi{10.1103/PhysicsPhysiqueFizika.1.195}.

\bibitem[Leggett and Garg(1985)]{LeggettGarg1985}
A.~J. Leggett and Anupam Garg.
\newblock Quantum mechanics versus macroscopic realism: {{Is}} the flux there when nobody looks?
\newblock \emph{Physical Review Letters}, 54\penalty0 (9):\penalty0 857--860, March 1985.
\newblock \doi{10.1103/PhysRevLett.54.857}.

\bibitem[Barrett and Kent(2004)]{BK2004}
Jonathan Barrett and Adrian Kent.
\newblock Non-contextuality, finite precision measurement and the {{Kochen}}--{{Specker}} theorem.
\newblock \emph{Studies in History and Philosophy of Science Part B: Studies in History and Philosophy of Modern Physics}, 35\penalty0 (2):\penalty0 151--176, June 2004.
\newblock ISSN 1355-2198.
\newblock \doi{10.1016/j.shpsb.2003.10.003}.

\bibitem[Kleinmann et~al.(2011)Kleinmann, G{\"u}hne, Portillo, Larsson, and Cabello]{Kleinmann_2011}
Matthias Kleinmann, Otfried G{\"u}hne, Jos{\'e}~R. Portillo, Jan-{\AA}ke Larsson, and Ad{\'a}n Cabello.
\newblock Memory cost of quantum contextuality.
\newblock \emph{New Journal of Physics}, 13\penalty0 (11):\penalty0 113011, November 2011.
\newblock ISSN 1367-2630.
\newblock \doi{10.1088/1367-2630/13/11/113011}.

\bibitem[Wilde and Mizel(2012)]{WildeMizel2012}
Mark~M. Wilde and Ari Mizel.
\newblock Addressing the {{Clumsiness Loophole}} in a {{Leggett-Garg Test}} of {{Macrorealism}}.
\newblock \emph{Foundations of Physics}, 42\penalty0 (2):\penalty0 256--265, February 2012.
\newblock ISSN 0015-9018, 1572-9516.
\newblock \doi{10.1007/s10701-011-9598-4}.

\bibitem[Kenfack and {\.Z}yczkowski(2004)]{KenfackZyczkowski2004}
Anatole Kenfack and Karol {\.Z}yczkowski.
\newblock Negativity of the {{Wigner}} function as an indicator of non-classicality.
\newblock \emph{Journal of Optics B: Quantum and Semiclassical Optics}, 6\penalty0 (10):\penalty0 396, August 2004.
\newblock ISSN 1464-4266.
\newblock \doi{10.1088/1464-4266/6/10/003}.

\bibitem[Davis et~al.(2021)Davis, Kumari, Mann, and Ghose]{Davis2021}
Jack Davis, Meenu Kumari, Robert~B. Mann, and Shohini Ghose.
\newblock Wigner negativity in spin- j systems.
\newblock \emph{Physical Review Research}, 3\penalty0 (3):\penalty0 033134, August 2021.
\newblock ISSN 2643-1564.
\newblock \doi{10.1103/PhysRevResearch.3.033134}.

\bibitem[Tsirelson(2006)]{Tsirelson2006}
Boris Tsirelson.
\newblock How often is the coordinate of a harmonic oscillator positive?
\newblock \penalty0 (arXiv:quant-ph/0611147), November 2006.
\newblock \doi{10.48550/arXiv.quant-ph/0611147}.

\bibitem[Zaw et~al.(2022)Zaw, Aw, Lasmar, and Scarani]{Zaw2022}
Lin~Htoo Zaw, Clive~Cenxin Aw, Zakarya Lasmar, and Valerio Scarani.
\newblock Detecting quantumness in uniform precessions.
\newblock \emph{Physical Review A}, 106\penalty0 (3):\penalty0 032222, September 2022.
\newblock \doi{10.1103/PhysRevA.106.032222}.

\bibitem[Aravinda et~al.(2017)Aravinda, Srikanth, and Pathak]{Aravinda2017}
S.~Aravinda, R.~Srikanth, and Anirban Pathak.
\newblock On the origin of nonclassicality in single systems.
\newblock \emph{Journal of Physics A: Mathematical and Theoretical}, 50\penalty0 (46):\penalty0 465303, November 2017.
\newblock ISSN 1751-8113, 1751-8121.
\newblock \doi{10.1088/1751-8121/aa8d29}.

\bibitem[Yu et~al.(2024)Yu, Wilhelm, Holmes, Vaartjes, Schwienbacher, Nurizzo, Kringh{\o}j, {van Blankenstein}, Jakob, Gupta, Hudson, Itoh, Murray, {Blume-Kohout}, Ladd, Dzurak, Sanders, Jamieson, and Morello]{Yu2024}
Xi~Yu, Benjamin Wilhelm, Danielle Holmes, Arjen Vaartjes, Daniel Schwienbacher, Martin Nurizzo, Anders Kringh{\o}j, Mark~R. {van Blankenstein}, Alexander~M. Jakob, Pragati Gupta, Fay~E. Hudson, Kohei~M. Itoh, Riley~J. Murray, Robin {Blume-Kohout}, Thaddeus~D. Ladd, Andrew~S. Dzurak, Barry~C. Sanders, David~N. Jamieson, and Andrea Morello.
\newblock Creation and manipulation of {{Schr{\"o}dinger}} cat states of a nuclear spin qudit in silicon.
\newblock \penalty0 (arXiv:2405.15494), May 2024.
\newblock \doi{10.48550/arXiv.2405.15494}.

\bibitem[Champion et~al.(2024)Champion, Wang, Parker, and Blok]{Champion2024}
Elizabeth Champion, Zihao Wang, Rayleigh Parker, and Machiel Blok.
\newblock Multi-frequency control and measurement of a spin-7/2 system encoded in a transmon qudit.
\newblock \penalty0 (arXiv:2405.15857), May 2024.
\newblock \doi{10.48550/arXiv.2405.15857}.

\bibitem[Roy et~al.(2024)Roy, Senanian, Wang, Wetherbee, Zhang, Cole, Larson, Yelton, Arora, McMahon, Plourde, Royer, and Fatemi]{Roy2024}
Saswata Roy, Alen Senanian, Christopher~S. Wang, Owen~C. Wetherbee, Luojia Zhang, B.~Cole, C.~P. Larson, E.~Yelton, Kartikeya Arora, Peter~L. McMahon, B.~L.~T. Plourde, Baptiste Royer, and Valla Fatemi.
\newblock Synthetic high angular momentum spin dynamics in a microwave oscillator.
\newblock \penalty0 (arXiv:2405.15695), September 2024.
\newblock \doi{10.48550/arXiv.2405.15695}.

\bibitem[Hofmann and Takeuchi(2004)]{Hofmann2004}
Holger~F. Hofmann and Shigeki Takeuchi.
\newblock Quantum-state tomography for spin-{$l$} systems.
\newblock \emph{Physical Review A}, 69\penalty0 (4):\penalty0 042108, April 2004.
\newblock \doi{10.1103/PhysRevA.69.042108}.

\bibitem[Zaw et~al.(2024)Zaw, Weilenmann, and Scarani]{Zaw2024}
Lin~Htoo Zaw, Mirjam Weilenmann, and Valerio Scarani.
\newblock A theory-independent bound saturated by quantum mechanics, January 2024.

\bibitem[Gupta et~al.(2024)Gupta, Vaartjes, Yu, Morello, and Sanders]{Gupta2024}
Pragati Gupta, Arjen Vaartjes, Xi~Yu, Andrea Morello, and Barry~C. Sanders.
\newblock Robust macroscopic {{Schr{\"o}dinger}}'s cat on a nucleus.
\newblock \emph{Physical Review Research}, 6\penalty0 (1):\penalty0 013101, January 2024.
\newblock ISSN 2643-1564.
\newblock \doi{10.1103/PhysRevResearch.6.013101}.

\bibitem[Asaad et~al.(2020)Asaad, Mourik, Joecker, Johnson, Baczewski, Firgau, M{\k a}dzik, Schmitt, Pla, Hudson, Itoh, McCallum, Dzurak, Laucht, and Morello]{Asaad2020}
Serwan Asaad, Vincent Mourik, Benjamin Joecker, Mark A.~I. Johnson, Andrew~D. Baczewski, Hannes~R. Firgau, Mateusz~T. M{\k a}dzik, Vivien Schmitt, Jarryd~J. Pla, Fay~E. Hudson, Kohei~M. Itoh, Jeffrey~C. McCallum, Andrew~S. Dzurak, Arne Laucht, and Andrea Morello.
\newblock Coherent electrical control of a single high-spin nucleus in silicon.
\newblock \emph{Nature}, 579\penalty0 (7798):\penalty0 205--209, March 2020.
\newblock ISSN 1476-4687.
\newblock \doi{10.1038/s41586-020-2057-7}.

\bibitem[{Fern{\'a}ndez de Fuentes} et~al.(2024){Fern{\'a}ndez de Fuentes}, Botzem, Johnson, Vaartjes, Asaad, Mourik, Hudson, Itoh, Johnson, Jakob, McCallum, Jamieson, Dzurak, and Morello]{Fernandez2024}
Irene {Fern{\'a}ndez de Fuentes}, Tim Botzem, Mark A.~I. Johnson, Arjen Vaartjes, Serwan Asaad, Vincent Mourik, Fay~E. Hudson, Kohei~M. Itoh, Brett~C. Johnson, Alexander~M. Jakob, Jeffrey~C. McCallum, David~N. Jamieson, Andrew~S. Dzurak, and Andrea Morello.
\newblock Navigating the 16-dimensional {{Hilbert}} space of a high-spin donor qudit with electric and magnetic fields.
\newblock \emph{Nature Communications}, 15\penalty0 (1):\penalty0 1380, February 2024.
\newblock ISSN 2041-1723.
\newblock \doi{10.1038/s41467-024-45368-y}.

\bibitem[Leuenberger and Loss(2003)]{Leuenberger2003}
Michael~N. Leuenberger and Daniel Loss.
\newblock Grover algorithm for large nuclear spins in semiconductors.
\newblock \emph{Physical Review B}, 68\penalty0 (16):\penalty0 165317, October 2003.
\newblock \doi{10.1103/PhysRevB.68.165317}.

\bibitem[Neeley et~al.(2009)Neeley, Ansmann, Bialczak, Hofheinz, Lucero, O'Connell, Sank, Wang, Wenner, Cleland, Geller, and Martinis]{Neeley2009}
Matthew Neeley, Markus Ansmann, Radoslaw~C. Bialczak, Max Hofheinz, Erik Lucero, Aaron~D. O'Connell, Daniel Sank, Haohua Wang, James Wenner, Andrew~N. Cleland, Michael~R. Geller, and John~M. Martinis.
\newblock Emulation of a {{Quantum Spin}} with a {{Superconducting Phase Qudit}}.
\newblock \emph{Science}, 325\penalty0 (5941):\penalty0 722--725, August 2009.
\newblock \doi{10.1126/science.1173440}.

\bibitem[Cybenko(2001)]{Cybenko2001}
George Cybenko.
\newblock Reducing quantum computations to elementary unitary operations.
\newblock \emph{Computing in Science \& Engineering}, 3:\penalty0 27--32, April 2001.
\newblock \doi{10.1109/5992.908999}.

\bibitem[Puri et~al.(2020)Puri, {St-Jean}, Gross, Grimm, Frattini, Iyer, Krishna, Touzard, Jiang, Blais, Flammia, and Girvin]{Puri2020}
Shruti Puri, Lucas {St-Jean}, Jonathan~A. Gross, Alexander Grimm, Nicholas~E. Frattini, Pavithran~S. Iyer, Anirudh Krishna, Steven Touzard, Liang Jiang, Alexandre Blais, Steven~T. Flammia, and S.~M. Girvin.
\newblock Bias-preserving gates with stabilized cat qubits.
\newblock \emph{Science Advances}, 6\penalty0 (34):\penalty0 eaay5901, August 2020.
\newblock \doi{10.1126/sciadv.aay5901}.

\bibitem[Gross(2021)]{Gross2021}
Jonathan~A. Gross.
\newblock Designing {{Codes}} around {{Interactions}}: {{The Case}} of a {{Spin}}.
\newblock \emph{Physical Review Letters}, 127\penalty0 (1):\penalty0 010504, July 2021.
\newblock \doi{10.1103/PhysRevLett.127.010504}.

\bibitem[Gross et~al.(2024)Gross, Godfrin, Blais, and {Dupont-Ferrier}]{Gross2024}
Jonathan~A. Gross, Cl{\'e}ment Godfrin, Alexandre Blais, and Eva {Dupont-Ferrier}.
\newblock Hardware-efficient error-correcting codes for large nuclear spins.
\newblock \emph{Physical Review Applied}, 22\penalty0 (1):\penalty0 014006, July 2024.
\newblock \doi{10.1103/PhysRevApplied.22.014006}.

\bibitem[Facon et~al.(2016)Facon, Dietsche, Grosso, Haroche, Raimond, Brune, and Gleyzes]{Facon2016}
Adrien Facon, Eva-Katharina Dietsche, Dorian Grosso, Serge Haroche, Jean-Michel Raimond, Michel Brune, and S{\'e}bastien Gleyzes.
\newblock A sensitive electrometer based on a {{Rydberg}} atom in a {{Schr{\"o}dinger-cat}} state.
\newblock \emph{Nature}, 535\penalty0 (7611):\penalty0 262--265, July 2016.
\newblock ISSN 1476-4687.
\newblock \doi{10.1038/nature18327}.

\bibitem[Chalopin et~al.(2018)Chalopin, Bouazza, Evrard, Makhalov, Dreon, Dalibard, Sidorenkov, and Nascimbene]{Chalopin2018}
Thomas Chalopin, Chayma Bouazza, Alexandre Evrard, Vasiliy Makhalov, Davide Dreon, Jean Dalibard, Leonid~A. Sidorenkov, and Sylvain Nascimbene.
\newblock Quantum-enhanced sensing using non-classical spin states of a highly magnetic atom.
\newblock \emph{Nature Communications}, 9\penalty0 (1):\penalty0 4955, November 2018.
\newblock ISSN 2041-1723.
\newblock \doi{10.1038/s41467-018-07433-1}.

\bibitem[Wheeler(1998)]{Wheeler1998}
Nicholas Wheeler.
\newblock Remarks concerning the status and some ramifications of {{Ehrenfest}}'s {{Theorem}}, 1998.
\newblock URL \url{https://www.reed.edu/physics/faculty/wheeler/documents/Quantum%20Mechanics/Miscellaneous%20Essays/Ehrenfest's%20Theorem.pdf}.

\bibitem[Hall(2013)]{hall2013quantum}
Brian~C. Hall.
\newblock \emph{Quantum {{Theory}} for {{Mathematicians}}}, volume 267 of \emph{Graduate {{Texts}} in {{Mathematics}}}.
\newblock Springer, New York, NY, 2013.
\newblock ISBN 978-1-4614-7115-8 978-1-4614-7116-5.
\newblock \doi{10.1007/978-1-4614-7116-5}.

\bibitem[Ringbauer et~al.(2015)Ringbauer, Duffus, Branciard, Cavalcanti, White, and Fedrizzi]{Ringbauer2015}
M.~Ringbauer, B.~Duffus, C.~Branciard, E.~G. Cavalcanti, A.~G. White, and A.~Fedrizzi.
\newblock Measurements on the reality of the wavefunction.
\newblock \emph{Nature Physics}, 11\penalty0 (3):\penalty0 249--254, March 2015.
\newblock ISSN 1745-2481.
\newblock \doi{10.1038/nphys3233}.

\bibitem[Fr{\"o}wis et~al.(2018)Fr{\"o}wis, Sekatski, D{\"u}r, Gisin, and Sangouard]{frowis2018macroscopic}
Florian Fr{\"o}wis, Pavel Sekatski, Wolfgang D{\"u}r, Nicolas Gisin, and Nicolas Sangouard.
\newblock Macroscopic quantum states: {{Measures}}, fragility, and implementations.
\newblock \emph{Reviews of Modern Physics}, 90\penalty0 (2):\penalty0 025004, May 2018.
\newblock \doi{10.1103/RevModPhys.90.025004}.

\end{thebibliography}

\begin{thebibliography}{9}
\providecommand{\natexlab}[1]{#1}
\providecommand{\url}[1]{\texttt{#1}}
\expandafter\ifx\csname urlstyle\endcsname\relax
  \providecommand{\doi}[1]{doi: #1}\else
  \providecommand{\doi}{doi: \begingroup \urlstyle{rm}\Url}\fi

\bibitem[Zaw et~al.(2022)Zaw, Aw, Lasmar, and Scarani]{Zaw2022}
Lin~Htoo Zaw, Clive~Cenxin Aw, Zakarya Lasmar, and Valerio Scarani.
\newblock Detecting quantumness in uniform precessions.
\newblock \emph{Physical Review A}, 106\penalty0 (3):\penalty0 032222, September 2022.
\newblock \doi{10.1103/PhysRevA.106.032222}.

\bibitem[Magnus(1985)]{eigenvalueDerivative}
Jan~R. Magnus.
\newblock On {{Differentiating Eigenvalues}} and {{Eigenvectors}}.
\newblock \emph{Econometric Theory}, 1\penalty0 (2):\penalty0 179--191, 1985.
\newblock ISSN 0266-4666.

\bibitem[Johansson et~al.(2013)Johansson, Nation, and Nori]{Johannson2013}
J.R. Johansson, P.D. Nation, and Franco Nori.
\newblock {{QuTiP}} 2: {{A Python}} framework for the dynamics of open quantum systems.
\newblock \emph{Computer Physics Communications}, 184\penalty0 (4):\penalty0 1234--1240, April 2013.
\newblock ISSN 00104655.
\newblock \doi{10.1016/j.cpc.2012.11.019}.

\bibitem[Agarwal(1981)]{Agarwal1981}
G.~S. Agarwal.
\newblock Relation between atomic coherent-state representation, state multipoles, and generalized phase-space distributions.
\newblock \emph{Physical Review A}, 24\penalty0 (6):\penalty0 2889--2896, December 1981.
\newblock \doi{10.1103/PhysRevA.24.2889}.

\bibitem[Morello et~al.(2010)Morello, Pla, Zwanenburg, Chan, Tan, Huebl, M{\"o}tt{\"o}nen, Nugroho, Yang, {van Donkelaar}, Alves, Jamieson, Escott, Hollenberg, Clark, and Dzurak]{Morello2010}
Andrea Morello, Jarryd~J. Pla, Floris~A. Zwanenburg, Kok~W. Chan, Kuan~Y. Tan, Hans Huebl, Mikko M{\"o}tt{\"o}nen, Christopher~D. Nugroho, Changyi Yang, Jessica~A. {van Donkelaar}, Andrew D.~C. Alves, David~N. Jamieson, Christopher~C. Escott, Lloyd C.~L. Hollenberg, Robert~G. Clark, and Andrew~S. Dzurak.
\newblock Single-shot readout of an electron spin in silicon.
\newblock \emph{Nature}, 467\penalty0 (7316):\penalty0 687--691, October 2010.
\newblock ISSN 1476-4687.
\newblock \doi{10.1038/nature09392}.

\bibitem[Pla et~al.(2013)Pla, Tan, Dehollain, Lim, Morton, Zwanenburg, Jamieson, Dzurak, and Morello]{Pla2013}
Jarryd~J. Pla, Kuan~Y. Tan, Juan~P. Dehollain, Wee~H. Lim, John J.~L. Morton, Floris~A. Zwanenburg, David~N. Jamieson, Andrew~S. Dzurak, and Andrea Morello.
\newblock High-fidelity readout and control of a nuclear spin qubit in silicon.
\newblock \emph{Nature}, 496\penalty0 (7445):\penalty0 334--338, April 2013.
\newblock ISSN 1476-4687.
\newblock \doi{10.1038/nature12011}.

\bibitem[Johnson et~al.(2022)Johnson, M{\k a}dzik, Hudson, Itoh, Jakob, Jamieson, Dzurak, and Morello]{Johnson2022}
Mark A.~I. Johnson, Mateusz~T. M{\k a}dzik, Fay~E. Hudson, Kohei~M. Itoh, Alexander~M. Jakob, David~N. Jamieson, Andrew Dzurak, and Andrea Morello.
\newblock Beating the {{Thermal Limit}} of {{Qubit Initialization}} with a {{Bayesian Maxwell}}'s {{Demon}}.
\newblock \emph{Physical Review X}, 12\penalty0 (4):\penalty0 041008, October 2022.
\newblock \doi{10.1103/PhysRevX.12.041008}.

\bibitem[Yu et~al.(2024)Yu, Wilhelm, Holmes, Vaartjes, Schwienbacher, Nurizzo, Kringh{\o}j, {van Blankenstein}, Jakob, Gupta, Hudson, Itoh, Murray, {Blume-Kohout}, Ladd, Dzurak, Sanders, Jamieson, and Morello]{Yu2024}
Xi~Yu, Benjamin Wilhelm, Danielle Holmes, Arjen Vaartjes, Daniel Schwienbacher, Martin Nurizzo, Anders Kringh{\o}j, Mark~R. {van Blankenstein}, Alexander~M. Jakob, Pragati Gupta, Fay~E. Hudson, Kohei~M. Itoh, Riley~J. Murray, Robin {Blume-Kohout}, Thaddeus~D. Ladd, Andrew~S. Dzurak, Barry~C. Sanders, David~N. Jamieson, and Andrea Morello.
\newblock Creation and manipulation of {{Schr{\"o}dinger}} cat states of a nuclear spin qudit in silicon.
\newblock \penalty0 (arXiv:2405.15494), May 2024.
\newblock \doi{10.48550/arXiv.2405.15494}.

\bibitem[McKay et~al.(2017)McKay, Wood, Sheldon, Chow, and Gambetta]{McKay2017}
David~C. McKay, Christopher~J. Wood, Sarah Sheldon, Jerry~M. Chow, and Jay~M. Gambetta.
\newblock Efficient {$Z$} gates for quantum computing.
\newblock \emph{Physical Review A}, 96\penalty0 (2):\penalty0 022330, August 2017.
\newblock \doi{10.1103/PhysRevA.96.022330}.

\end{thebibliography}

\begingroup
\renewcommand{\addcontentsline}[3]{} 

\let\oldaddcontentsline\addcontentsline
\renewcommand{\addcontentsline}[3]{}
\providecommand{\noopsort}[1]{}\providecommand{\singleletter}[1]{#1}%

\endgroup

\newpage
\
\newpage

\beginsupplement

\title{Supplementary Information: Certifying the quantumness of a nuclear spin qudit through its uniform precession}

\maketitle
\onecolumngrid
\tableofcontents
    
\section{Theory}

\subsection{Classical bounds for the even and uneven protocols}
In the experiment, we performed the precession protocol with $K$ angles, where the angles were not always equally spaced. Label the measured angles as $\varphi_1 \leq \varphi_2 \leq \dots \leq \varphi_{K}$ such that $\forall k : \varphi_{k}-\varphi_1 \leq 2\pi$. That is, we place all the angles in a circle and label them in increasing order in a clockwise direction. As long as we have chosen the angles such that
\begin{equation}\label{eq:condition}
\forall k:\pqty{\varphi_{1+\pqty{k-1+\frac{K-1}{2}}\bmod K}-\varphi_k}\bmod 2\pi \leq \pi,
\end{equation}
the classical bound is $\mathbf{P}^c(\{\varphi_k\}_{k=1}^{K}) = (1+1/K)/2$.

This can be proven geometrically: for any initial state $(x_0,y_0)$, the $K$ points will take angles $\varphi_0 + \varphi_k$ with $\varphi_0 := \atan(y_0,x_0)$. Let $\varphi_0 + \varphi_{k_0}$ be the point in the $x >0 \land y<0$ plane closest to the $y$ axis.  Since $[(\varphi_0 + \varphi_{1+(k_0-1+\frac{K-1}{2})\bmod K})-(\varphi_0 + \varphi_{k_0})]\bmod 2\pi \leq \pi$, this means that there must be at least $(K+1)/2$ points on the left of a straight line from  $\varphi_0 + \varphi_{k_0}$ to the origin, which means that there must be at least $(K-1)/2$ points on the negative $x$ plane, which therefore implies that there must be at most $(K+1)/2$ points on the positive $x$ plane. Since this is true for every $\varphi_0$, Eq.~\eqref{eq:condition} implies that $P^c(\{\varphi_k\}_{k=1}^{K}) \leq \mathbf{P}^c(\{\varphi_k\}_{k=1}^{K}) = (1+1/K)/2$.

\subsection{Analytical expressions for the quantum protocol}
For a given initial state $\rho$ of a spin-$J$ system, where we will only consider dimensions $d=2J+1$ that are even, the expected score after performing the precession protocol is
\begin{equation}
    P^d(\{\varphi_k\}_{k=1}^K) = \frac{1}{K}\sum_{k=1}^K\tr[e^{i\varphi_k \hat{I}_z} \rho e^{-i\varphi_k \hat{I}_z}  \operatorname{Pos}(\hat{I}_x)]
    = \tr[\rho Q^d(\{\varphi_k\})],
\end{equation}
where $Q^d(\{\varphi_k\}_{k=1}^K) = \frac{1}{K}\sum_{k=1}^K e^{-i\varphi_k \hat{I}_z}  \operatorname{Pos}(\hat{I}_x)e^{i\varphi_k \hat{I}_z}$. Its matrix elements are known to be \cite{Zaw2022}
\begin{equation}
    \begin{aligned}
    \bra{m}Q^d(\{\varphi_k\}_{k=1}^K)\ket{m'}
    &= \frac{1}{2}\Bigg\{
        \delta_{m,m'}  + \frac{
            (-1)^{\frac{m'-m-1}{2}} 2^{-(2J-1)}
        }{m'-m}\sqrt{
            \left(\begin{matrix}
                 2\Big\lfloor\frac{J+m}{2}\Big\rfloor \\
                 \Big\lfloor\frac{J+m}{2}\Big\rfloor & 
            \end{matrix}\right)
            \left(\begin{matrix}
                 2\Big\lfloor\frac{J-m}{2}\Big\rfloor \\
                 \Big\lfloor\frac{J-m}{2}\Big\rfloor & 
            \end{matrix}\right)
            \left(\begin{matrix}
                 2\Big\lfloor\frac{J+m'}{2}\Big\rfloor \\
                 \Big\lfloor\frac{J+m'}{2}\Big\rfloor & 
            \end{matrix}\right)
            \left(\begin{matrix}
                 2\Big\lfloor\frac{J-m'}{2}\Big\rfloor \\
                 \Big\lfloor\frac{J-m'}{2}\Big\rfloor & 
            \end{matrix}\right)
        } \\
    &\qquad{}\times{}\sqrt{
        (J+m)^{(J+m)\bmod 2}
        (J-m)^{(J-m)\bmod 2}
        (J+m')^{(J+m')\bmod 2}
        (J-m')^{(J-m')\bmod 2}
    }\\
    &\qquad{}\times{}\frac{1}{K}\sum_{k=1}^K e^{-i\varphi_k(m-m')} 
    \Bigg\}.
    \end{aligned}
\end{equation}
As such, for a fixed set of angles $\{\varphi_k\}_{k=1}^K$, we can find the maximum quantum violation
\begin{equation}
\mathbf{P}^d(\{\varphi_k\}_{k=1}^K) := \max_\rho \tr[\rho Q^d(\{\varphi_k\}_{k=1}^K)]
\end{equation}
by solving for the eigenvalues of $Q^d(\{\varphi_k\}_{k=1}^K)$ with standard numerical tools.

For the special case that $K \geq 3$, $J=K/2$, and $\varphi_k = \varphi_0+2\pi k/K$, this gives
\begin{equation}\label{eq:Q^{K+1}}
    Q^{K+1}(\{\varphi_0 + 2\pi k/K\}_{k=1}^K) = \frac{1}{2}\mathbb{I} + (-1)^{\frac{K-1}{2}}
    2^{-K} \left(\begin{matrix}
        K-1\\
        \frac{K-1}{2}
    \end{matrix}\right)
    \left(
        e^{-i\varphi_0 K}\ketbra{\tfrac{K}{2}}{-\tfrac{K}{2}} +
        e^{i\varphi_0 K}\ketbra{-\tfrac{K}{2}}{\tfrac{K}{2}}
    \right).
\end{equation}
The maximum quantum score in this case is therefore $\mathbf{P}_K^{K+1} = \frac{1}{2} + 2^{-K} \left(\begin{matrix}K-1\\\frac{K-1}{2}\end{matrix}\right)$, which is achieved by the state
\begin{equation}
    \frac{1}{\sqrt{2}}\left(\ket{-\tfrac{K}{2}} + (-1)^{\frac{K-1}{2}} e^{-i\varphi_0 K}\ket{\tfrac{K}{2}}\right).
\end{equation}
Hence, given the initial state $\ket{\rm cat_{K/2}} = \frac{1}{\sqrt{2}}(\ket{-\tfrac{K}{2}} + (-1)^{\frac{K-1}{2}}\ket{\tfrac{K}{2}})$, the expected score is
\begin{equation}
    P_K^{K+1} = \frac{1}{2} +  2^{-K} \left(\begin{matrix}K-1\\\frac{K-1}{2}\end{matrix}\right) \cos(\varphi_0 K).
\end{equation}
The values of $\mathbf{P}_{K}^{d \leq 3K}$ and their corresponding states can be found using similar steps to be \cite{Zaw2022}
\begin{equation}
\begin{aligned}
\mathbf{P}_K^{K < d \leq 3K} &= 
\frac{1}{2}\left(
1 + \frac{2^{-(2J-1)}}{\cos\lambda}\sqrt{
    \left(\begin{matrix}
        K-1\\
        \frac{K-1}{2}
    \end{matrix}\right)
    \left(\frac{2J}{K} - (d \bmod 2)\right)
    \left(\begin{matrix}
        2\lfloor J \rfloor\\
        \lfloor J \rfloor
    \end{matrix}\right)
    \left(\begin{matrix}
        2\lfloor J - \frac{K}{2} \rfloor\\
        \lfloor J - \frac{K}{2} \rfloor
    \end{matrix}\right)
}
\right) \\
\ket{\mathbf{P}_K^{K < d \leq 3K}} &= \frac{1}{\sqrt{2}}\ket{-J+K} + (-1)^{\frac{K-1}{2}}\frac{1}{\sqrt{2}}\times\begin{cases}
    \ket{-J} & \text{if $K < d \leq 2K$,} \\
    \cos\lambda\ket{-J} + \sin\lambda\ket{-J+2K} & \text{if $2K < d \leq 3K$.}
\end{cases},
\end{aligned}
\end{equation}
where 
\begin{equation}
\lambda = \begin{cases}
    0 & \text{if $K < d \leq 2K$,}\\
    \arctan(\sqrt{\frac{
        [J-(d\bmod2)K]\left(\scriptsize\begin{matrix}
            2\lfloor J-K \rfloor \\
            \lfloor J-K \rfloor
        \end{matrix}\right)
    }{
        [J-K]\left(\scriptsize\begin{matrix}
            2\lfloor J-K \rfloor \\
            \lfloor J-K \rfloor
        \end{matrix}\right)\left(\scriptsize\begin{matrix}
            2K \\
            K
        \end{matrix}\right)
    }}) & \text{if $2K < d \leq 3K$.}
\end{cases}
\end{equation}

\subsection{Fidelity bounds for the cat state}
If we perform the evenly-spaced precession protocol on a spin-$J$ particle with $J=K/2$, the score $P^{K+1}_K = \tr[\rho Q^{K+1}(\{2\pi k/K\}_{k=1}^K)]$ is achieved, where $\rho$ is the initial state prepared in the experiment. Let $\ket{\pm \rm cat_{K/2}} := \frac{1}{\sqrt{2}}(\ket{-\tfrac{K}{2}} \pm (-1)^{\frac{K-1}{2}}\ket{\tfrac{K}{2}})$. We are interested in the fidelity $\mathcal{F} = \bra{+ \rm cat_{K/2}}\rho\ket{+ \rm cat_{K/2}}$ against the cat state $\ket{+ \rm cat_{K/2}}$. A large observed value of $P^{K+1}_K$ must imply a large value of $\mathcal{F}$, and we can indeed bound the latter with the former.

From Eq.~\eqref{eq:Q^{K+1}},
\begin{equation}
\begin{aligned}
    P^{K+1}_K &= \tr[\rho Q^{K+1}(\{2\pi k/K\}_{k=1}^K)] \\
    &= \frac{1}{2} + 
    2^{-K} \left(\begin{matrix}
        K-1\\
        \frac{K-1}{2}
    \end{matrix}\right)
    \left(
        \bra{+\rm cat_{K/2}}\rho\ket{+\rm cat_{K/2}} - \bra{-\rm cat_{K/2}}\rho\ket{-\rm cat_{K/2}}
    \right) \\
    &\geq \frac{1}{2} + 2^{-K}\left(\begin{matrix}
        K-1\\
        \frac{K-1}{2}
    \end{matrix}\right)\mathcal{F}.
\end{aligned}
\end{equation}
After rearranging the terms, we obtain the lower bound
\begin{equation}
    \mathcal{F} \geq \frac{2P^{K+1}_K-1}{2^{-(K-1)}\left(\begin{matrix}
        K-1\\
        \frac{K-1}{2}
    \end{matrix}\right)}.
\end{equation}
In particular for $K=7$, this is $\mathcal{F} \geq 16(2P^8_7-1)/5$.

\subsection{Optimization of the quantum score for the uneven protocol}
Given some set of angles $\Phi_K := \{\varphi_k\}_{k=1}^{K}$, the maximum eigenvalue for a fixed $\Phi_K$ is given by
\begin{equation}
    \mathbf{P}^d(\Phi_K) = \bra{\mathbf{P}^d(\Phi_K)}Q^d(\Phi_K) \ket{\mathbf{P}^d(\Phi_K)},
\end{equation}
where $Q^d(\{\varphi_k\}_{k=1}^K) = \frac{1}{K}\sum_{k=1}^K e^{-i\varphi_k \hat{I}_z}  \operatorname{Pos}(\hat{I}_x)e^{i\varphi_k \hat{I}_z}$ as before and $| \mathbf{P}^d(\Phi_K) \rangle$ is the eigenstate of $Q^d(\{\varphi_k\}_{k=1}^K) $ corresponding to its maximum eigenvalue.

For a given number of angles $K$, we wish to find as large a violation of the classical bound as possible. To do so, we start with some initial set of angles $\Phi_K$, then improve the violation using the gradient \cite{eigenvalueDerivative}
\begin{equation}
\begin{aligned}
    \frac{\partial\mathbf{P}(\Phi_K)}{\partial\varphi_k} &=
    \frac{1}{K} \frac{\partial}{\partial \varphi_k}  \bra{\mathbf{P}^d(\Phi_K)} \operatorname{Pos}(\cos\varphi_k \hat{I}_x + \sin\varphi_k \hat{I}_y) \ket{\mathbf{P}^d(\Phi_K)} \\
    &= \frac{i}{K} \bra{\mathbf{P}^d(\Phi_K)} e^{-i\varphi_k \hat{I}_z} \comm{\operatorname{sgn} \hat{I}_x}{\hat{I}_z} e^{i\varphi_k \hat{I}_z} \ket{\mathbf{P}^d(\Phi_K)} \\
    &= i\frac{2J+1}{2K} \bra{\mathbf{P}^d(\Phi_K)} e^{-i\theta_k \hat{I}_z} e^{-i\frac{\pi}{2} \hat{I}_y} \pqty{
        \ketbra{-\tfrac{1}{2}}{\tfrac{1}{2}} -
        \ketbra{\tfrac{1}{2}}{-\tfrac{1}{2}} 
    } e^{i\frac{\pi}{2} \hat{I}_y} e^{i\varphi_k \hat{I}_z} \ket{\mathbf{P}^d(\Phi_K)} .
\end{aligned}
\end{equation}
Here, the closed-form expression for $[\operatorname{sgn}\hat{I}_x,\hat{I}_z]$ was previously worked out in Ref.~\cite{Zaw2022}. The reported maximally-violating states were found numerically by performing gradient descent on several initial choices of $\Phi_K$ sampled over the probing angles that satisfy Eq.~\eqref{eq:condition}.

\subsection{Visualization with the Wigner function}

For visualisation of the spin states in the main text, we plot the spherical spin-Wigner function from Qutip \cite{Johannson2013}, defined as:  

\begin{equation}
\label{spin Wigner 1}
    W(\theta, \varphi) = \sqrt{\frac{2}{\pi}}\ 
    \sum_{k=0}^{2I}\sum_{q=-k}^{k} Y_{kq}(\theta,\varphi)\rho_{kq} \
\end{equation}
where $\theta$ and $\varphi$ are the polar and azimuthal angles of the Bloch sphere, $I$ is the length of the spin vector ($I=7/2$ for the full $d=8$ spin), $Y_{kq}$ are the spherical harmonics and $\rho_{kq}$ is an element of the density matrix decomposed in the spherical harmonic basis \cite{Agarwal1981}. \\\\
\noindent We use a truncated density matrix for visualizing the Wigner spheres of $d<8$ subspaces (for instance in Fig.4 of the main text). The truncation discards the density matrix elements outside the subspace, and renormalizes the values such that $\mathrm{Tr}(\rho)$=1. Note that this truncation is for visualization purposes only. It does not affect the estimate of the fidelity from quantum state tomography in Sec.~\ref{sec:density_matrix_tomo}. For those calculations, the entire density matrix is taken into account.

\section{Experimental details}

\subsection{Readout}
\label{sec:readout}
We read out the nuclear spin state by employing an ancillary electron coupled to a single electron transistor (SET) as a mediator to perform approximate quantum non-demolition (QND) measurements \cite{Morello2010, Pla2013}. 
During a nuclear readout operation, we perform the following steps: 
\begin{enumerate}
    \item \textbf{Load a $\ket{\downarrow}$ electron onto the $\Sb$ nucleus.}

    This step neutralizes the donor and changes the Hamiltonian to: 

    \begin{equation}
    \hat{\mathcal{H}}_{D^0} = \gamma_{\rm{e}}B_{0}\hat{S_{z}} - \gamma_\mathrm{n} B_{0}\hat{I_{z}} + A \cdot \hat{\textbf {S}}\cdot \hat{\textbf{I}} + \hat{\mathcal{H}}_{Q^0},
    \label{neutral Hamiltonian}
    \end{equation}

    This Hamiltonian gives rise to 8 electron spin resonance (ESR) frequencies, approximately separated by the hyperfine interaction $A \approx 98$~MHz. The other terms in the Hamiltonian are: $\gamma_e, \gamma_n$, the gyromagnetic ratios of the electron and the nucleus, $B_0$, the static magnetic field, $\hat{S}_z$ and $\hat{I}_z$, the electron and nuclear spin-z operators. Furthermore, $\hat{\textbf {S}}\cdot \hat{\textbf{I}}$ is the inner product between the electron and nuclear spin vectors and $\hat{\mathcal{H}}_{Q^0}$ denotes the quadrupole interaction in the neutral donor.

    We ensure the loaded electron is in the $\ket{\downarrow}$ state by an initialization process known as ``Maxwell's demon'' \cite{Johnson2022, Yu2024}.
    
    \item \textbf{Perform selective adiabatic electron spin resonance (aESR) to flip the electron spin conditional on the nuclear spin state. }

    This constitutes a bit flip on the electron, conditional on the nuclear spin eigenstate $\ket{m_z}$.
    
    \item \textbf{Wait a time $\tau_\mathrm{read}$.}
    
    If the electron state has been flipped to $\ket{\uparrow}$, it is allowed to tunnel onto the SET, producing an electrical current. If we measure a current, we label the electron as $\ket{\uparrow}$. Otherwise, it is labeled as $\ket{\downarrow}$
    
    \item \textbf{Repeat steps 1, 2 and 3 several times for each nuclear spin state.}

    In this work, we typically used $N_{\rm read} = 10$. After completing the readout steps, we assign a $\ket{\uparrow}$-proportion to each nuclear spin state.

    \item \textbf{Assign the most likely nuclear spin state $\ket{m_z}$ by a majority vote.}

    The nuclear spin state is identified as state state with the maximum $\ket{\uparrow}$-proportion.

\end{enumerate}  

\begin{figure}[h]
    \centering
    \includegraphics[width=\textwidth]{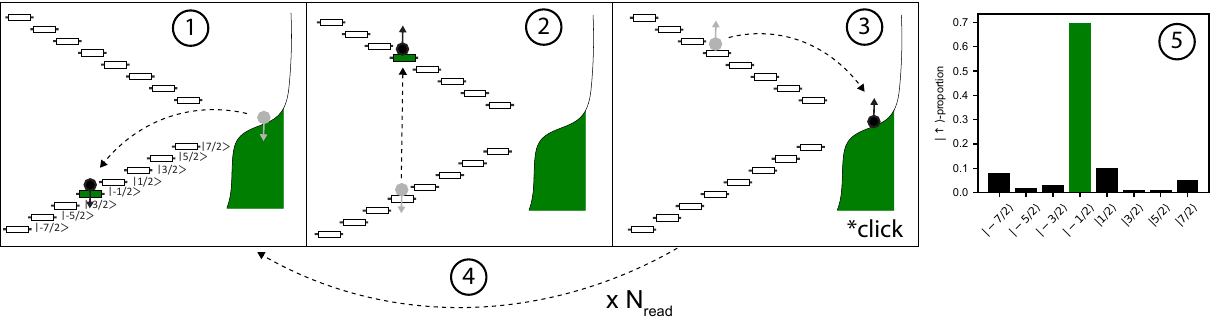}
    \caption{\textbf{The five steps for nuclear readout}}
    \label{fig:readout}
\end{figure}
\FloatBarrier

\subsection{Arbitrary state initialization}
The realization of the protocol requires the initialization of spin states $\ket{\psi}  =\sum_0^7 c_m \ket{m-7/2}$ where the coefficients $c_m$ have arbitrary phases and amplitudes.
To do so we start by preparing the spin in the ground state by repeating the two following steps until the spin state is projected in $\ket{-7/2}$ at step 1 :
\begin{enumerate}
    \item Read out the nuclear spin state (see section \ref{sec:readout}) and obtain output $\ket{m-7/2}$, $m \in \{0, ..., 7\}$.
    \item Perform sequential $R_{\text{-}y}(\pi)$ pulses in between state $\ket{m-7/2}$ and $\ket{(m-1)-7/2}$, $\ket{(m-1)-7/2}$ and $\ket{(m-2)-7/2}$, etc...
\end{enumerate}

\begin{figure}
    \centering
    \includegraphics[width=0.8\textwidth]{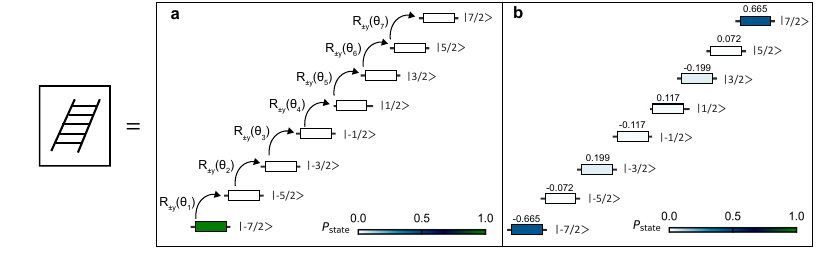}
    \caption{\textbf{Arbitrary nuclear state initialization via ladder operations.}\textbf{a} The nuclear spin is initialized in the ground state $\ket{-7/2}$. We then apply 7 sequential $R_{\pm y}(\theta_k)$ to distribute the state amplitude across the 8 eigenstates of the nuclei.
    \textbf{b} Example of the outcome of the initialization protocol for the $\ket{\mathbf{P^8_{\Tilde{5}}}}$ state.}
    \label{fig:arbitrary-init}
\end{figure}

At the end of this procedure the nucleus is initialized in $\ket{-7/2}$ with fidelity close to $99 \%$ limited by the QND readout fidelity \cite{Yu2024}.
From there we distribute the amplitude of the state across the 8 coefficients $c_k$ using a laddering operation. As depicted in Fig.~\ref{fig:arbitrary-init}, we perform 7 sequential rotations around either $\text{+}y$ or $\text{-} y$ in between adjacent eigenstates with angle $\theta_m$. The rotation angle is calculated such that rotation $m$ will leave the targeted amplitude $c_m$ at each eigenstate and is computed as follows:    

\begin{equation}
\theta_m = \cos^{-1} \left( \frac{\abs{c_m}}{\sqrt{\sum_{i=m}^{7}c_i^2}} \right)
\label{eq:angle-laddering}
\end{equation}

To determine along which axis the rotation must be performed we notice that for all the states that need to be initialized in this article the phase difference in between two consecutive coefficients of $\ket{\psi}$ is either $0$ or $\pi$ (see example in \ref{fig:arbitrary-init}).
Therefore during the laddering we can choose during pulse $m$ to transfer the amplitude with the same phase as $c_m$ to $c_{m+1}$ using a $R_{\text{+}y}(\theta_m)$ rotation or add a phase $\pi$ to $c_{m+1}$ using a $R_{\text{-}y}(\theta_m)$ rotation.

To summarise we adapt the rotation angle of each sequential pulse depending on the amplitude to transfer during the ladder operation. Each rotation is performed around the $\text{+} y$, $\text{-} y$ axis of the Bloch sphere if the two successive coefficients have the same, opposite sign respectively.

\subsection{Uniform precession using virtual SNAP gates}
\label{sec:snap}

\noindent In our implementation of the uniform precession protocol, we generate virtual $R_z(\varphi)$ rotations using SNAP (selective number-dependent arbitrary phase) gates. This technique, widely used in superconducting qubit systems \cite{McKay2017}, has recently been adapted for the spin-7/2 qudit system \cite{Yu2024}. 

\noindent To describe the $R_z-$rotation for a spin-7/2 system, i.e. the uniform precession around the $I_z$-axis, we define first:
\begin{equation}
    \hat{I}_{z} = \sum_{m=0}^{7} \ket{m-7/2}\bra{m-7/2}.
\end{equation}
Rotations around the $\hat{I}_z$ axis are described by the Hamiltonian $\hat{\mathcal{H}}_{\rm precession} = -\omega \hat{I}_z$, which gives rise to the $R_z$ unitary:
\begin{equation}
    R_z(\varphi) = e^{-i \varphi \hat{I}_z} =  \sum_{m=0}^{7} e^{-im \varphi} \ket{m-7/2}\bra{m-7/2}
\end{equation}
where $\omega$ is an angular frequency and $\varphi = \omega t$ represents the azimuthal angle on the Bloch sphere. The challenge is to generate this unitary using the virtual SNAP gates, as defined in equation~\ref{eq:snap_unitary}.
\\\\
The SNAP gate is a diagonal unitary of arbitrary phases, defined as: 

\begin{equation}
    \hat{\mathcal{U}}_{\rm{SNAP}}
     \left( \Vec{\xi} \right) = \sum_{m=0}^{7} e^{i\xi_m} \ket{m-7/2}\bra{m-7/2}
     \label{eq:snap_unitary}
\end{equation}
where $\vec{\xi} = \{\xi_0, ... , \xi_7\}$. It becomes clear that we can see the $R_z-$rotation as a ``linear SNAP'' gate; a special case of the virtual SNAP gate, where all phases on the diagonal have a linear relationship proportional to the $\hat{I}_z$ operator. \\\\

\begin{figure}
    \centering
    \includegraphics{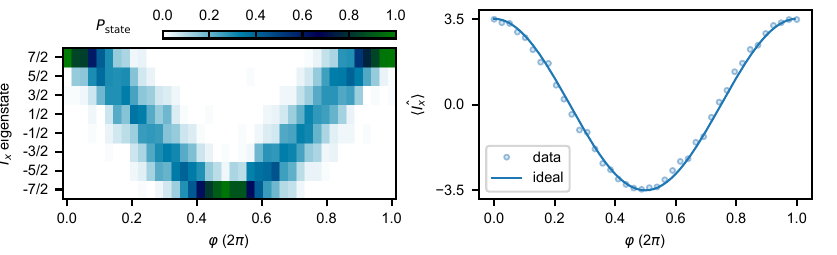}
    \caption{\textbf{$R_z$ rotation through virtual SNAP gates.} An initial spin-coherent state oriented along the $I_x$-direction undergoes a rotation around $I_z$ through a virtual snap gate with angle $\varphi$. The resulting state is measured in the $I_x$-eigenbasis by performing a basis rotation (see main text Fig.3)}
    \label{fig:rz_rotation}
\end{figure}

\noindent In the spin-7/2 system the phases of the eight $I_z$-eigenstates in a spin-7/2 system are controlled by seven clocks within the FPGA hardware. By shifting the clock phases $\phi_i$, we can create virtual SNAP gates without performing a physical rotation on the qudit. The $I_z$-axis rotation $R_z$ then becomes manifest due to a phase shift in the subsequent pulse. \\\\
The virtual phases $\xi_m$ relate to the physical phase shifts $\Delta\phi_i$ as a cumulative sum:
\begin{equation}
    \xi_m = - \sum_{i=1}^m{\Delta \phi_i},
    \label{phase_update}
\end{equation}
 To generate $R_z(\varphi)$, we uniformly shift all physical phases by an equal amount $\varphi$. This results in $\xi_m = -m\varphi$, leading to: 

\begin{equation}
         \left( \Delta\phi_i = \varphi ~:~ \forall i \right) \rightarrow
    \hat{\mathcal{U}}_{\rm{SNAP}}
= e^{i7\varphi/2}\hat{R}_z({\varphi}),
\end{equation}
where we can ignore the global phase factor $e^{i7\varphi/2}$.\\\\
\noindent Thus, by uniformly shifting the phases of subsequent rotations by $\varphi$, we effectively construct the $R_z(\varphi)$ rotation using virtual SNAP gates. An experimental demonstration of the virtual $R_z$ rotation is provided in Fig.~\ref{fig:rz_rotation}

\subsection{Subspace rotations}

The protocol is performed in various subspaces of the 8-dimensional Hilbert space, for instance, in Fig.4 of the main text we present the protocol outcomes for cat states in $d=8,6$ and $d=4$. For the protocol within a subspace, we need to generalize two operations to a $d-$dimensional subspace:
\begin{enumerate}
    \item The precession around the z-axis denoted by $R^d_z(\varphi)$.

    We provide an example for $d=4$ below. The subspace is spanned by $\{\ket{-3/2}, \ket{-1/2},\ket{1/2},\ket{3/2}\}$. The $R^4_z(\varphi)$ rotation is defined as:
\begin{equation}
    R^4_z(\varphi) = \sum_{m=2}^5 e^{-i m \varphi} \ket{m-7/2}\bra{m-7/2}.
\end{equation}
This unitary is simply achieved by shifting only the physical phases within the subspace by an equal amount $\varphi$.

    \item The SU(2)-covariant rotation around the y-axis to perform the basis rotation, denoted by $R^d_{-y}(\pi/2)$

    This rotation for $d=4$ is defined as: 

\renewcommand{\arraystretch}{}
\begin{equation}
    R^4_{-y}(\pi/2) =  \frac{1}{2}
\left(
\begin{array}{cccccccc}
0 & 0 & 0 & 0 & 0 & 0 & 0 & 0 \\
0 & 0 & 0 & 0 & 0 & 0 & 0 & 0 \\
0 & 0 & 0 & \sqrt{3} & 0 & 0 & 0 & 0 \\
0 & 0 & \sqrt{3} & 0 & 2 & 0  & 0 & 0 \\
0 & 0 & 0 & 2 & 0 & \sqrt{3}  & 0 & 0 \\
0 & 0 & 0 & 0 & \sqrt{3} & 0 & 0 & 0 \\
0 & 0 & 0 & 0 & 0 & 0 & 0 & 0 \\
0 & 0 & 0 & 0 & 0 & 0 & 0 & 0 \\
\end{array}
\right).
\label{eq:subspace_drive}
\end{equation}

\end{enumerate}

The subspace rotation is achieved by rescaling the drive amplitudes (equation 2 in the main text) to match the matrix in equation~\ref{eq:subspace_drive}.

\subsection{Device characterization}

\begin{figure}[h]
    \centering
    \includegraphics{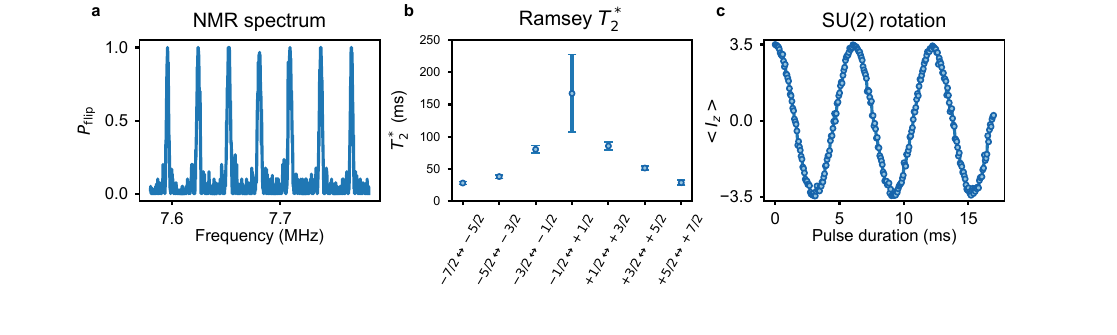}
    \caption{\textbf{Device characterization. } \textbf{a} NMR spectrum, showing 7 transitions separated by an quadrupole splitting of $\langle f_q \rangle=28.3$~kHz. \textbf{b} Corresponding $T_2*$ decay times af adjacent nuclear spin levels extracted with a Ramsey experiment. \textbf{c} SU(2) rotation around the -$I_y$-axis showing a $\pi$-time of 3 ms.}
    \label{fig:characterization}
\end{figure}

The $\Sb$ implanted device we used to perform the protocol is identical to the one in Ref.\cite{Yu2024}, where a thorough characterization can be found. In Tab.~\ref{tab:char} and Fig.~\ref{fig:characterization} we provide a summary of the characterization results.

\renewcommand{\arraystretch}{1.2}
\begin{table}[h]
\begin{tabular}{l|l|l|l}
Transition              & $f_{\rm NMR}$ (MHz) & $\pi$ time (ms) & $T_2^*$ (ms) \\ \hline
-7/2 $\rightarrow$ -5/2 & 7.5963           & 0.328           & 27.77    \\
-5/2 $\rightarrow$ -3/2 & 7.6246           & 0.254           & 37.79    \\
-3/2 $\rightarrow$ -1/2 & 7.6529           & 0.227           & 80.18    \\
-1/2 $\rightarrow$ 1/2  & 7.6812           & 0.220           & 167.17   \\
1/2 $\rightarrow$ 3/2   & 7.7095           & 0.223           & 85.43    \\
3/2 $\rightarrow$ 5/2   & 7.7739           & 0.251           & 51.09    \\
5/2 $\rightarrow$ 7/2   & 7.7664           & 0.329           & 28.37 

\end{tabular}
\caption{Experimental data for rotations in qubit subspaces \cite{Yu2024}. $f_{\rm NMR}$ indicates the NMR frequency and $T_2^*$ the Ramsey coherence time. }
\label{tab:char}  
\end{table}. 

\FloatBarrier
\section{Additional datasets}
\subsection{Reconstructed density matrices}
\label{sec:density_matrix_tomo}

\begin{figure}
    \centering
    \includegraphics{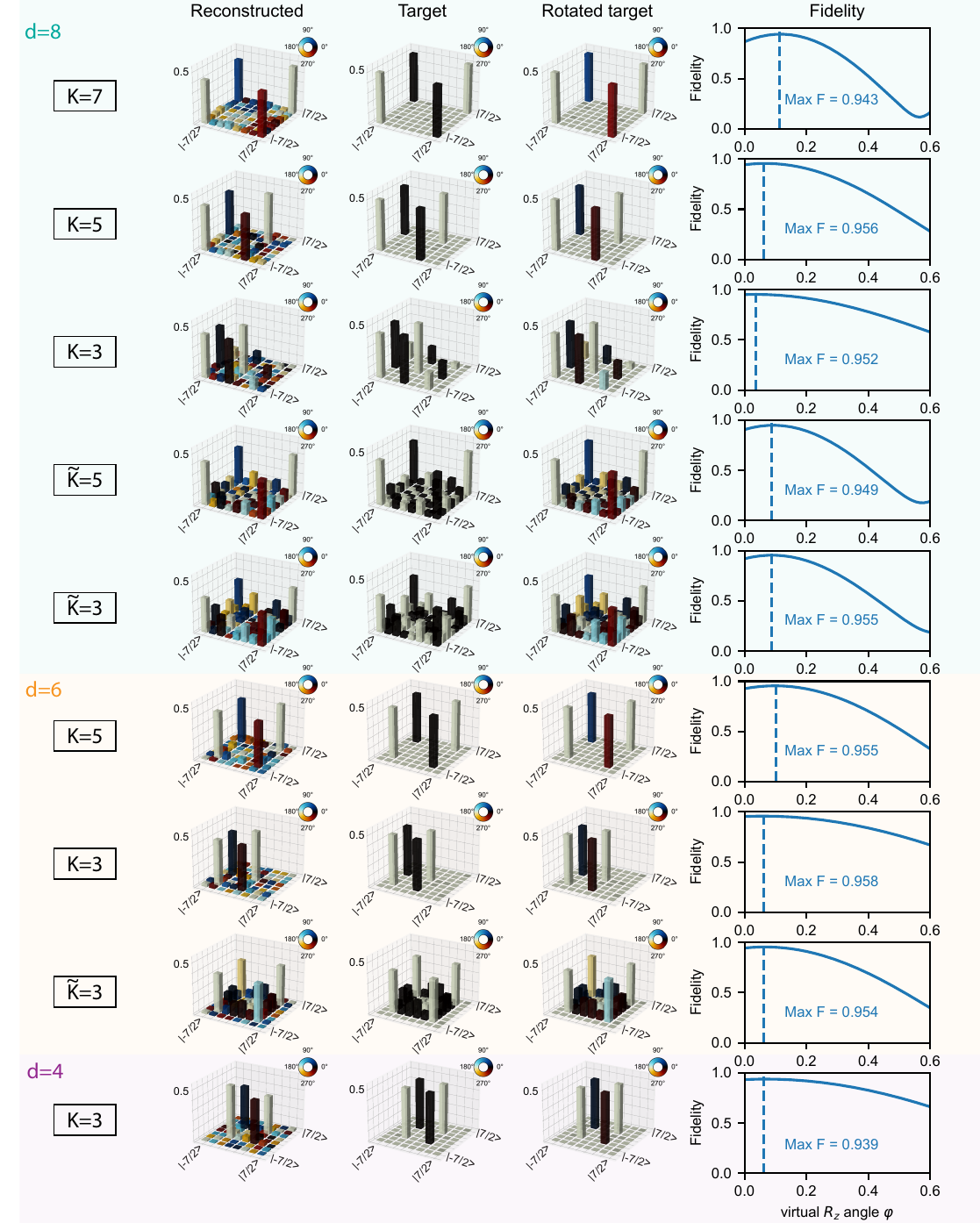}
    \caption{Reconstructed density matrices and fidelities}
    \label{fig:sup_density_matrix}
\end{figure}

We performed quantum state tomography on all states that violate the classical bound displayed in Tab.1 of the main text. The tomographic measurement protocol is outlined in Supplementary section 7 in Ref.~\cite{Yu2024}. The results are shown in Fig.~\ref{fig:sup_density_matrix}, where we display the reconstructed density matrix and the target density matrix. \\\\
The quantum state tomography allows us to extract the fidelity between the intended target state and the state that was actually prepared.  It also allows us to diagnose why that fidelity is less than 1.  Notably, the reconstructed states systematically display a phase shift, which we can attribute to an AC stark shift due to off-resonant driving of the next-nearest neighbour transition in the ladder initialization process. This phase shift is deterministic, and can be partially corrected by a virtual $R_z$ rotation. For the protocol results, such a phase shift does not matter: the quantum score is invariant under $R_z$ rotations as we sweep over the entire range of $R_z$ angles $\varphi$. 

\subsection{Error bars on fidelity}

\begin{figure}
    \centering
    \includegraphics[width=\textwidth]{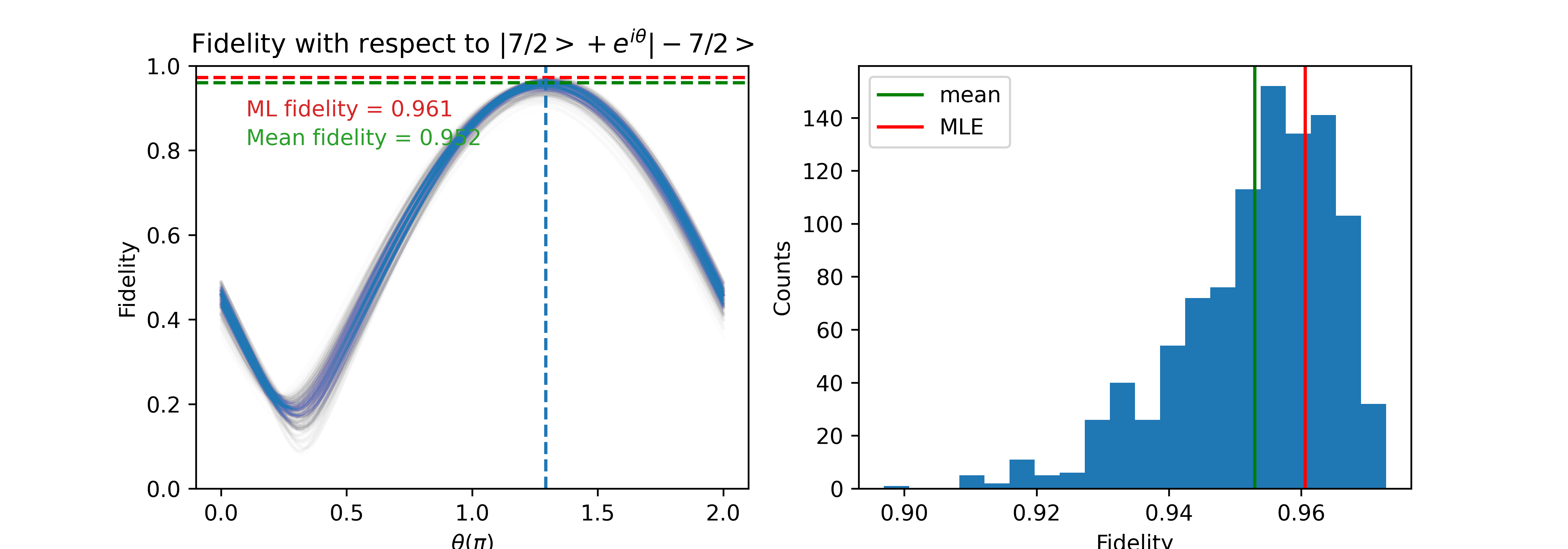}
    \caption{Estimation of the error on the fidelity}
    \label{fig:fidelity_error}
\end{figure}

To estimate error bars on the fidelity, we perform a bootstrapping method with a sample size of N=1000. Each of the N resampled datasets is prepared by randomly resampling the state probabilities for each measurement setting with replacement, yielding a vector of state probabilities that is multinomially distributed with a mean equal to the actual observed data. For each bootstrap sample, we compute the most likely density matrix, and compute the fidelity. The errorbar on the fidelity is then estimated as the standard deviation over the bootstrapped samples. Figure \ref{fig:fidelity_error} shows all N=1000 instances of the fidelity calculation as a function of the phase $\theta$, together with a histogram of the maximum fidelities. The mean fidelity (the value quoted in the main text of the paper), indicated in green in the figure, is close to the maximum likelihood estimate of the aggregate bootstrapping data. We observe a slightly skewed distribution, which can likely be ascribed to the upper limit of 1 of the fidelity. 

\FloatBarrier
\subsection{Protocol data and comparison with simulation for d=8,6,4 and 2.}

\noindent In this section, we display the results of the protocol measurements for all states in Tab.1 of the main text. For each state we show the state populations of the $I_x$-eigenstates, positivity operator $\langle {\rm Pos}(\hat{I}_x) \rangle$ and the measured quantum score $P_K^d$ as a function of the precession angle $\varphi$. \\\\
\noindent Additionally, we compare $\langle {\rm Pos}(\hat{I}_x) \rangle$ to simulated protocol data. We simulated the protocol both with the ideal density matrix, i.e., the one that maximizes the quantum score, as well as the reconstructed density matrix (Fig.~\ref{fig:sup_density_matrix}). We observe good agreement of the measured data and simulated data, and the gap between the measured quantum score and the ideal quantum score can be partially explained by the non-unity fidelity of the initial state. \\\\
\noindent Furthermore, we included two examples of semi-classical states that do not break the classical bound. Fig.~\ref{fig:scs} shows protocol data for the $d=8$ spin coherent state, and demonstrate that this semi-classical state state does not violate the classical bound for any choice of $K$ ($K=3$ and $K=7$ are shown). Lastly, in Fig.~\ref{fig:d2} we show an example of a case where $K>d$, which theoretically results in a quantum score of $P_K^d=\frac{1}{2}$ \cite{Zaw2022}. Here, we prepared the superposition state $\frac{1}{\sqrt{2}}\left( \ket{-1/2}-\ket{1/2}\right)$, which is also a spin-coherent state in the $d=2$ subspace. Indeed, we show that the measured $P_3^2$ is close to $\frac{1}{2}$.

\begin{figure}
    \centering
    \includegraphics{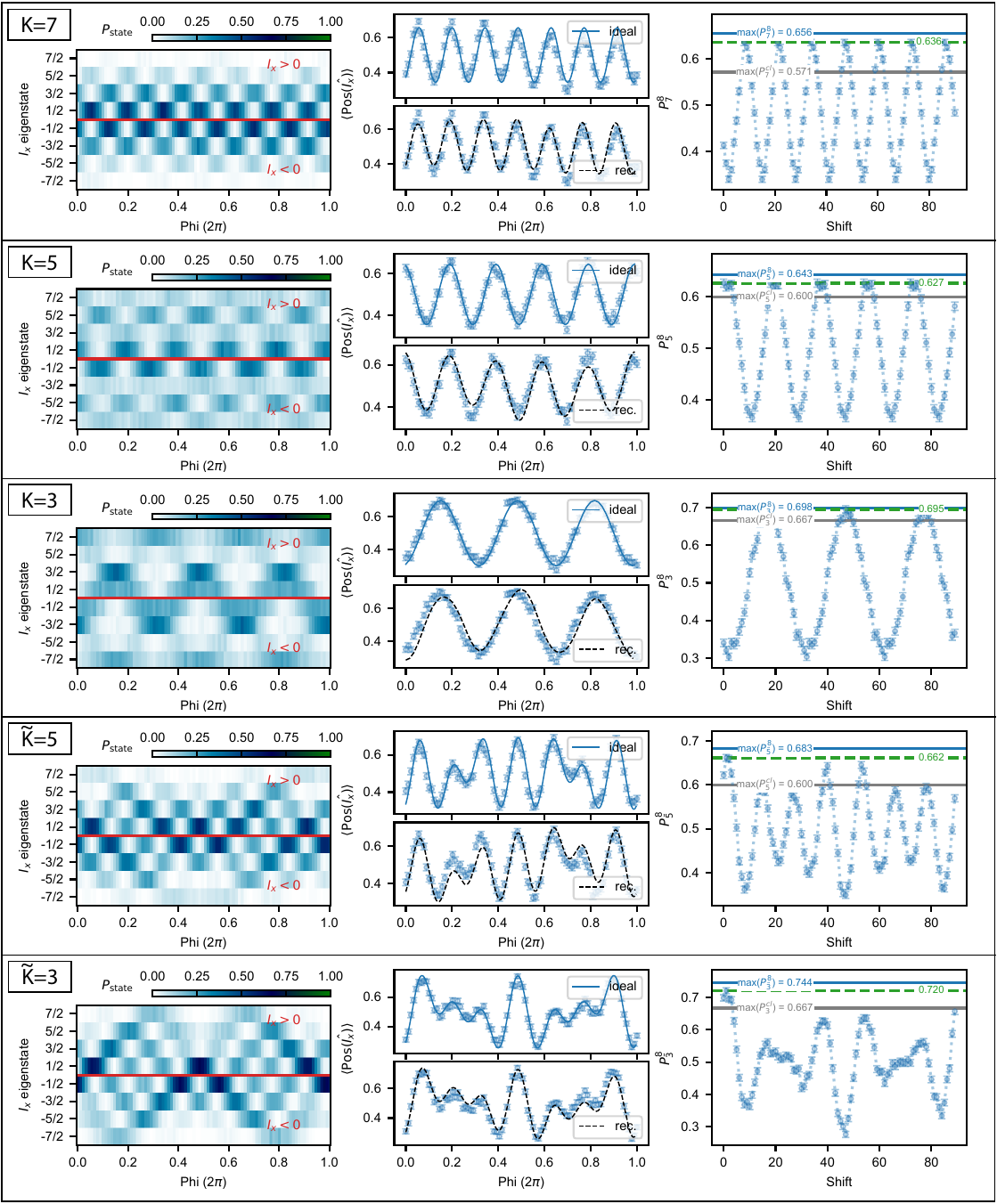}
    \caption{Precession protocol data and comparison with simulation for $d=8$}
    \label{fig:d8}
\end{figure}

\begin{figure}
    \centering
    \includegraphics{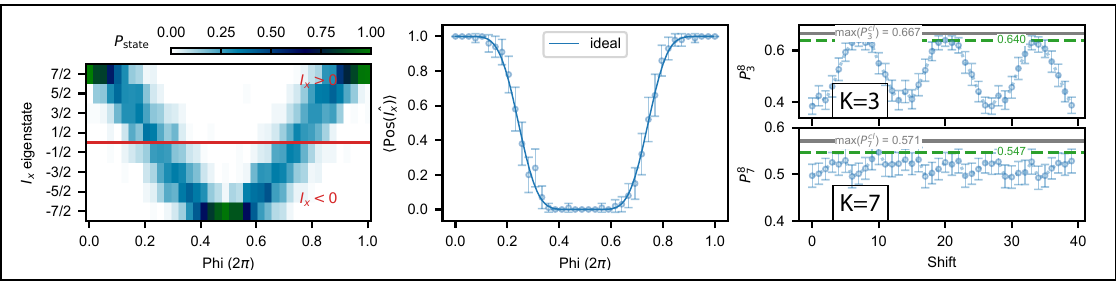}
    \caption{Precession protocol data and comparison with simulation for a $d=8$ spin coherent state}
    \label{fig:scs}
\end{figure}

\begin{figure}
    \centering
    \includegraphics{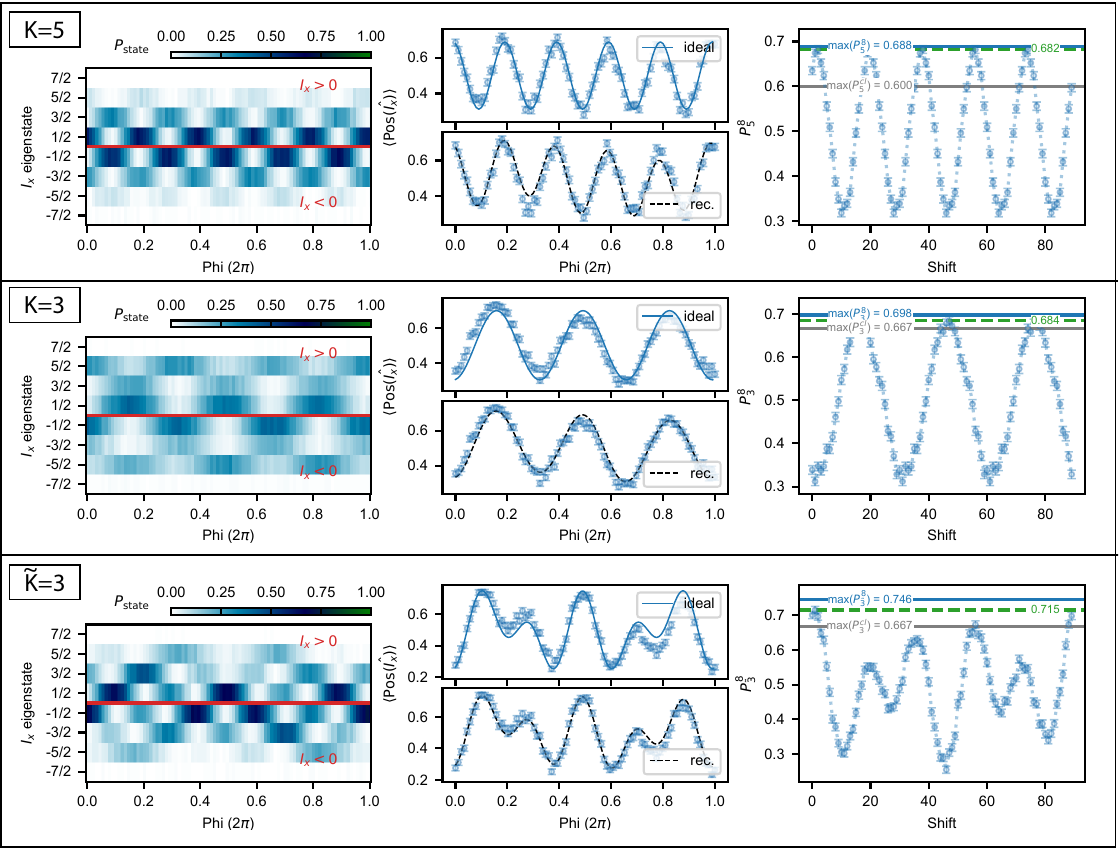}
    \caption{Precession protocol data and comparison with simulation for $d=6$}
    \label{fig:d6}
\end{figure}

\begin{figure}
    \centering
    \includegraphics{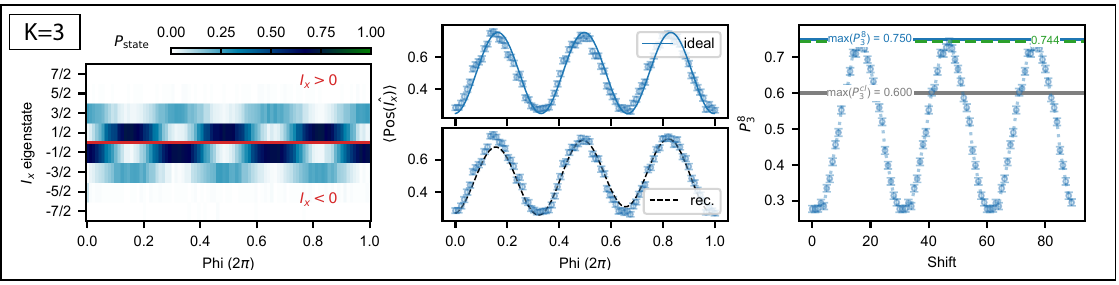}
    \caption{Precession protocol data and comparison with simulation for $d=4$}
    \label{fig:d4}
\end{figure}

\begin{figure}
    \centering
    \includegraphics{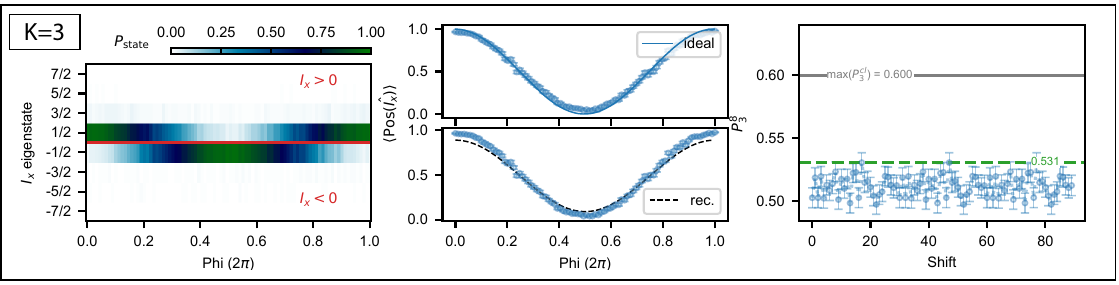}
    \caption{Precession protocol data and comparison with simulation for $d=2$}
    \label{fig:d2}
\end{figure}

\FloatBarrier

\providecommand{\noopsort}[1]{}\providecommand{\singleletter}[1]{#1}%


\end{document}